\documentclass[journal,onecolumn,draftclsnofoot,12pt]{IEEEtran}

\usepackage{pdfsync}
\usepackage{wrapfig}
\usepackage{multirow}
\usepackage{amsfonts}
\usepackage{amssymb}
\usepackage{amsmath}
\usepackage{graphicx}
\usepackage{pdfpages}
\usepackage{cite}
\usepackage[T1]{fontenc}
\usepackage{balance}
\usepackage{algorithm}
\usepackage{array}
\usepackage{algpseudocode}
\usepackage{float}
\usepackage{amsmath,amsthm,amssymb}
\usepackage{amsmath,amsthm}
\usepackage{bbm}

\usepackage{booktabs}
\usepackage{threeparttable}
\usepackage{subfigure}
\newtheorem{lem}{Theorem}
\newcommand{\bs}{\boldsymbol}

\theoremstyle{lemma}

\begin{document}

\title{Massive Access in Multi-cell Wireless Networks Using Reed-Muller Codes}

\author{Pei~Yang,
        Dongning~Guo,
       and Hongwen Yang%
\thanks{Pei~Yang is with School of Wireless Communication Center, Beijing University of Posts and Telecommunications, Beijing 100876, China and also with the Department of Electrical and Computer Engineering, Northwestern University, Evanston, IL 60208 USA (e-mail: yp@bupt.edu.cn).}%
\thanks{Dongning~Guo is with the Department of Electrical and Computer Engineering, Northwestern University, Evanston, IL 60208 USA (e-mail: dguo@northwestern.edu).}
\thanks{Hongwen~Yang is with School of Wireless Communication Center, Beijing University of Posts and Telecommunications, Beijing 100876, China (e-mail: yanghong@bupt.edu.cn).}%
}

\maketitle

\begin{abstract}
\boldmath
Providing connectivity to a massive number of devices is a key challenge in 5G wireless systems.
In particular, it is crucial to develop efficient methods for active device identification and message decoding in a multi-cell network with fading and path loss uncertainties.
In this paper, we design such a scheme using second-order Reed-Muller (RM) sequences.
For given positive integer $m$, a codebook is generated with up to $2^{m(m+3)/2}$ codewords of length $2^m$, where each codeword is a unique RM sequence determined by a matrix-vector pair with binary entries.
This allows every device to send $m(m+3)/2$ bits of information where an arbitrary number of these bits can be used to represent the identity of a node, and the remaining bits represent a message.
There can be up to $2^{m(m+3)/2}$ devices in total.
Using an iterative algorithm, an access point can estimate the matrix-vector pairs of each nearby device, as long as not too many devices transmit simultaneously.
To improve the performance, we also describe an enhanced RM coding scheme with slotting.
We show that both the computational complexity and the error performance of the latter algorithm exceed another state-of-the-art algorithm.
The device identification and message decoding scheme developed in this work can serve as the basis for grant-free massive access for billions of devices with hundreds of simultaneously active devices in each cell.

\end{abstract}

\begin{IEEEkeywords}
Channel estimation, many access, neighbor discovery, Reed-Muller code, device identification.
\end{IEEEkeywords}

\section{Introduction}
One of the promises of 5G wireless communication systems is to support large scale machine-type communication (MTC), that is, to provide connectivity to a massive number of devices as in the Internet of Things \cite{Durisi,Tullberg,Larsson}.
A key characteristic of MTC is that the user traffic is typically sporadic so that in any given time interval, only a small fraction of devices are active.
Also, short packets are the most common form of traffic generated by sensors and devices in MTC.
This requires a fundamentally different design than that for supporting sustained high-rate mobile broadband communication.

To extend classical information theory to massive access, an information-theoretic paradigm called {\em many-user access} was proposed by Chen, Chen, and Guo in \cite{Chen,Chen2}, where the number of users grows with the coding blocklength.
It was shown to be asymptotically optimal for active users to simultaneously transmit their identifying signatures followed by their message bearing codewords.
Meanwhile, massive access signaling techniques have received a lot of attention in the past few years; see examples \cite{Yu,Yu2,Yu3,Yu4,Hanzo,Zhang,Andreev,Polyanskiy,Poor,Ravi,Yener,Letaief,Sohrabi,Ahn,Lau,Jia,Robert,ZhangLi,Applebaum} and references therein.

To discovery and identify the active users, each user must be assigned a unique sequence.
Since there are an enormous number of potential users in the system, the signature sequences of most users cannot be orthogonal to each other, so multiaccess interference is unavoidable.
The key challenge is to design a large sequence space with a reliable detection method.
In \cite{Kim}, the author proposed a random access scheme based on the Zadoff-Chu (ZC) sequence, which allows relatively large number of preambles.
However, the collision rate for contention-based random access \cite{Hasan,Carvalho} is still too high in the context of massive access.
We note that if the number of users to support is in the millions or more, we must design a highly structured codebook.
This is because it is computationally infeasible to visit all user codewords in one frame slot.
To address this, Reed-Muller (RM) sequence-based massive access schemes were proposed in \cite{Luo,Zhang,Robert,Howard,ZhangLi,Hanzo}.
At typical code lengths, the number of RM sequences is several orders of magnitude larger than the number of ZC sequences.
In fact the code book size can be so large that every user is assigned a different signature in any practical system.
In addition, the RM sequences are well structured to allow very fast detection algorithms.
In particular, an asynchronous random access algorithm based on convex optimization was presented in \cite{Applebaum}, which overcomes the issues of contention-based methods.
A structured group sparsity estimation method for device identification and channel estimation was proposed in \cite{Letaief}.
Reference \cite{Ahn} proposed an expectation propagation-based joint active user detection and channel estimation technique, and the active device detection for a cloud-radio access network was studied in \cite{Lau}.
To lower the complexity for massive user detection, a dimension reduction-based joint active user detection and channel estimation method was proposed in \cite{Jia}.

Due to the sporadic traffic in MTC, the active user detection problem is usually cast as a compressed sensing problem.
Typical compressed sensing algorithms can be applied for detecting active users and channel estimation \cite{Yu,Yu2,Yu3,Yu4,Sohrabi,Schepker,Gil,Giannakis,Jeong,Mir,Shim}.
With known user channel coefficients, several compressed sensing methods is first applied for active user detection \cite{Jeong,Mir,Shim}. Then joint active user detection and channel estimation methods for single-cell massive connectivity scenario via approximate message passing (AMP) were proposed in \cite{Sohrabi,Yu,Yu2,Yu3}.
The authors in \cite{Yu4} further extended the network model to multi-cell massive multiple-input-multiple-output (MIMO) and cooperative MIMO systems.
In addition, greedy compressed sensing algorithm was designed for sparse signal recovery based on orthogonal matching pursuit \cite{Schepker,Gil}.
We note that algorithms using random codes and/or AMP type of decoding do not scale to millions of potential devices.

Different from \cite{Chen,Chen2}, the author in \cite{Polyanskiy2} advocates the unsourced model focusing on how to recover messages from nodes using the same codebook.
Identification can still be accomplished by letting all or part of the payload describe the identity.
In this case, coded slotted ALOHA schemes were proposed in \cite{Paolini,Casini}, where each message is sent over multiple slots, and information is passed between slots to recover messages lost due to collision.
The authors of \cite{Ordentlich} proposed the $T$-fold ALOHA scheme such that the decoder can jointly decode all the messages if at most $T$ messages are sent in the same slot, whereas nothing is decoded when more than $T$ messages are transmitted within the same slot.
And the performance of the $T$-fold ALOHA scheme is further improved in \cite{Vem}, and \cite{Kowshik} proposed the $T$-ALOHA based random access scheme for handing Rayleigh fading channels.
Following the framework in \cite{Polyanskiy2}, an enhanced chirp reconstruction algorithm based on RM sequences was proposed in \cite{Robert}, assuming identical known channel gain for all users.
It is shown in \cite{Robert} that the worst-case complexity is sub-linear in the number of codewords, which makes it an attractive algorithm for message decoding in MTC.

In this paper, we consider the problem of joint device identification/decoding and channel estimation.
We first derive an unambiguous relationship between the RM sequence and its subsequences, where each RM sequence is uniquely determined by a matrix-vector pair.
Given this relationship, an iterative RM sequence detection and channel estimation algorithm is proposed.
To enhance the performance of the RM detection algorithm, we further divide the codeword into multiple slots and transmit each user's message in exactly two of the multiple slots \cite{Robert}.
If a message is successfully decoded in one of the two transmitted slots, the decoded messages are then propagated to the other slots that the messages are transmitted, which can significantly improve the successful decoding probability.

The main differences between our algorithms and the state-of-the-art RM detection algorithms are: 1) The algorithm in \cite{Robert} works only for the additive white Gaussian noise (AWGN) channel (the channel estimation problem is thus not considered therein), and the authors in \cite{Hanzo} considered only the small-scale fading. In contrast, we consider both small-scale fading and large-scale fading in a general network setting; 2) The algorithm in \cite{Hanzo} requires the receiver to know the number of active devices in the cell, and its value has a great impact on the decoding result. In addition, the performance of the algorithm in \cite{Hanzo} degrades dramatically when the number of active devices is large. In contrast, by adopting slotting and message passing, our algorithm performs gracefully as the number of active devices increases.
The main contributions are summarized as follows:
\begin{itemize}
\item We first derive an explicit relationship between an RM sequence and its subsequences. Based on this relationship, we propose an iterative active device detection and channel estimation algorithm, Algorithm 1, for fading channels with path loss, where each device's matrix-vector pair and the channel coefficient can be estimated successively.
\item Inspired by \cite{Robert}, we further describe an enhanced RM coding scheme with slotting. The corresponding detection algorithm is referred to as Algorithm 2.
\item The computational complexity and performance of Algorithm 1 are comparable to the RM decoding algorithm in \cite{Hanzo}, while the computational complexity and performance of Algorithm 2 are notably improved, which makes it one important step closer to a practical algorithm.
\item While many papers in the literature study massive access, this work (alone with \cite{Hanzo}) is one of the few that can truly accommodate billions of devices and more.
\end{itemize}

The remainder of this paper is organized as follows.
The system model is presented in Section \ref{Secsystemmodel}.
Section \ref{RMrelationship} outlines the relationship between the RM sequence and its subsequences, which is the basis of the RM decoding algorithm.
The device identification and channel estimation algorithm is expressed in Section \ref{useridentification}, while the enhanced RM decoding algorithm utilizing slotting and message passing is explained in Section \ref{algorithm2}.
Further, the computation complexity analysis is given in Section \ref{CCA}.
Section \ref{numericlresult} presents the numerical results, while Section \ref{conclusion} concludes the paper.

Throughout the paper, boldface uppercase letters stand for matrices while boldface lowercase letters represent column vectors.
The superscripts $(\cdot)^{\rm T}$, $(\cdot)^{*}$, and $(\cdot)^\dag$ denote the transpose, complex conjugate, and conjugate transpose operator, respectively.
The complex number field is denoted by $\mathbb C$.
$\|\bs x\|$ denotes the 2-norm of a vector $\bs x$ and $|A|$ denotes the cardinality of set $A$.
$\bs I_n$ denotes an $n\times n$ identity matrix.
$\oplus$ denotes element-wise modulo-2 addition.

\section{System Model}\label{Secsystemmodel}
\subsection{Channel Model}
We consider a multi-cell network over a large region. User devices are randomly distributed in a similar fashion as was proposed in \cite{Guo} (for an ad hoc network therein). Let time be slotted. A device may or may not be active in a slot. It is assumed that active devices transmit simultaneously. Each device sends $B$ bits, which consists of either an identity or a message or a combination of both.
Let $\bs\Phi=\{Z_k\}$ denote the set of active devices on the plane, where $Z_k$ represents the location of a device.
Without loss of generality, we focus on one access point (AP) and assume that the AP is located at the origin of the plane.
Let $K=|\bs\Phi|$ represents the (random) number of active devices on the plane where $K\ll 2^B$, and denote the index set of the active devices as $ \{1,2,\cdots,K\}$ without loss of generality.
Given this, the AP's received signal is written as
\begin{align}\label{systemmodel}
    \bs y^m=\sqrt{\gamma}\sum_{k=1}^Kh_k\bs c_k^m+{\bs z}^m
\end{align}
where $\bs c_k^m\in{\mathbb C}^{n}$ is the codeword corresponding to device $k$ with $n=2^m$ being the codeword length; $h_k$ denotes the complex channel coefficient between device $k$ and the AP; ${\bs z}^m\sim{\cal CN}(\bs 0,\bs I_n)$ denotes the circularly symmetric complex AWGN; $\gamma$ denotes the transmit power and we define $\gamma|h_k|^2$ as the nominal received signal-to-noise ratio (SNR) for device $k$.

We further divide the active device set $\bs \Phi$ into in-cell device set and out-of-the-cell device set of the AP according to the nominal SNR between the AP and the devices. In other words, if the nominal SNR of a device to the AP is larger than a threshold (to be determined in Section \ref{PM}), then this device is considered an in-cell device of the AP.
Without loss of generality, denote the index set of the in-cell devices as $\{1,2,\cdots,K^{*}\}$ and denote the index set of the out-of-the-cell devices as $ \{K^{*}+1,\cdots,K\}$.
In this case, \eqref{systemmodel} can be further expressed as
\begin{equation}
\begin{split}\label{systemmodel2}
    \bs y^m&=\sqrt{\gamma}\sum_{k=1}^{K^{*}}h_k\bs c_k^m + \sqrt{\gamma}\sum_{k=K^{*}+1}^Kh_k\bs c_k^m +\bs z^m.\\
\end{split}
\end{equation}
The purpose of the AP is to identify all in-cell devices and/or decode their messages, where transmissions from out-of-the-cell devices are regarded as interference.

Our approach is to assign each message a length-$2^m$ second-order RM sequence $\bs c^m$ determined by a symmetric binary matrix $\bs P^m\in{\mathbb Z}_2^{m\times m}$ and a binary vector $\bs b^m\in{\mathbb Z}_2^{m}$.
Since $\bs P^m$ is determined by $\frac12m(m+1)$ bits and $\bs b^m$ is determined by $m$ bits, each sequence encodes $\frac12m(m+3)$ bits.
Given the matrix-vector pair $(\bs P^m,\bs b^m)$, the $j$-th entry of the RM sequence $\bs c^m$ can be written as \cite{Robert}
\begin{equation}
\begin{split}\label{cm2}
    c^m_j=i^{ 2(\bs b^m)^{\rm T}\bs a_{j-1}^m+(\bs a_{j-1}^m)^{\rm T}\bs P^m\bs a_{j-1}^m }
\end{split}
\end{equation}
where $i^2=-1$, $\bs a_{j-1}^m$ is the $m$-bit binary expression of $(j-1)$. Eq. \eqref{cm2} indicates that $c^m_j\in \{1,-1,i,-i\}$.

As a toy example, in the special case of $m=2$, there are altogether 32 RM sequences of length 4. This entire set of sequences is enumerated in Table \ref{ExampleRMcodes}.
In numerical results in Section \ref{numericlresult}, we simulate systems with $m$ as high as 14, which accommodate up to $2^{119}$ sequences of length 16,384.

\begin{table*}[htbp]
\renewcommand\arraystretch{0.8}
\centering
\caption{The entire set of 2nd-order RM sequences in the special case of $m=2$.}
\label{ExampleRMcodes}
\setlength{\tabcolsep}{1.2mm}{
\begin{tabular}{rrrr rrrr rrrr rrrr rrrr rrrr rrrr rrrr }
  \hline
1&1&1&1&1&1&1&1& 1&1&1&1&1&1&1&1& 1&1&1&1&1&1&1&1& 1&1&1&1&1&1&1&1 \\
1&1&$i$&$i$& 1&1&$i$&$i$& -1&-1&-$i$&-$i$&-1&-1&-$i$&-$i$& 1&1&$i$&$i$& 1&1&$i$&$i$& -1&-1&-$i$&-$i$&-1&-1&-$i$&-$i$\\
1&1&1&1&$i$&$i$&$i$&$i$& 1&1&1&1&$i$&$i$&$i$&$i$&  -1&-1&-1&-1&-$i$&-$i$&-$i$&-$i$& -1&-1&-1&-1&-$i$&-$i$&-$i$&-$i$ \\
1&-1&-$i$&$i$&$i$&-$i$&-1&1& -1&1&$i$&-$i$&-$i$&$i$&1&-1& -1&1&$i$&-$i$&-$i$&$i$&1&-1& 1&-1&-$i$&$i$&$i$&-$i$&-1&1\\
  \hline
\end{tabular}
}
\end{table*}

\subsection{Propagation Model and Cell Coverage}\label{PM}
Consider a multiaccess channel with devices distributed across the plane according to a homogeneous Poisson point process with
intensity $\lambda$. The number of devices in a region with its area equal to $S$ is a Poisson random variable with mean $\lambda S$.

The small-scale fading between device and the AP is modeled by an independent Rayleigh random variable with unit mean.
The large-scale fading is modeled by the free-space path loss which attenuates over distance with some path loss exponent $\alpha>2$.

Let $R_{k}$ and $G_{k}$ denotes the distance and the small scale Rayleigh fading gain between device $k, k=1,2,\cdots,K$, and the AP, respectively.
Then the channel gain is expressed as
\begin{equation}
\begin{split}
    |h_k|^2=R_k^{-\alpha}G_k,
\end{split}
\end{equation}
where the phase of $h_k$ is uniformly distributed on $[0,2\pi)$.

The coverage of the AP can be defined in many different ways. According to \cite{Guo}, device $k$ and the AP are neighbors of each other if the channel gain exceeds a certain threshold $\theta$.
Assume device $k$ and the AP are neighbors, i.e., $R_k^{-\alpha}G_k>\theta$, we have $R_k<\left({\frac{G_k}{\theta}}\right)^{1/{\alpha}}$.
Under the assumption that all devices form a p.p.p., for given $G_k$, device $k$ is uniformly distributed in a disk centered at the AP with radius $\left({\frac{G_k}{\theta}}\right)^{1/{\alpha}}$.
According to \cite{Guo}, the distribution of $|h_k|$ is
\begin{equation}
\begin{split}
    p\left(|h_k|\right)=\left\{\begin{array}{l}
                          \frac{4}{\alpha}\frac{  \theta^{{2}/{\alpha}}  }  { |h_k|^{{4}/{\alpha}+1} }, \quad |h_k|>\sqrt{\theta}\\
                          0, \qquad\quad\quad\ \ {\rm otherwise}
                        \end{array}\right.
\end{split},
\end{equation}
and the average number of neighbors of the AP is
\begin{equation}
\begin{split}\label{averageneighbor}
    K^{*}=\frac2{\alpha}\pi\lambda\theta^{-2/\alpha}\Gamma\left(\frac2{\alpha}\right)
\end{split},
\end{equation}
where $\Gamma(\cdot)$ is the Gamma function.
In addition, the sum power of all out-of-the-cell devices is \cite{Guo}
\begin{equation}
\begin{split}\label{nonneighbor}
    \sigma^2=\frac{4}{\alpha(\alpha-2)}\pi\lambda\gamma\theta^{1-2/\alpha}\Gamma\left(\frac{2}{\alpha}\right)
\end{split}.
\end{equation}

\section{A Property of RM Sequences}\label{RMrelationship}
In this section, we introduce a fundamental property of RM sequences, which underlies the following detection algorithm.

let $m$ be a given positive number. Let $\bs b^s=[b^m_1,b_2^m,\cdots,b_s^m]^{\rm T}$ be a binary $s$-tuple. For $s=2,\cdots,m$, we have
\begin{equation}
\begin{split}\label{b}
    \bs b^s=\left[\begin{array}{c}
               \bs b^{s-1} \\
               b_s^m
             \end{array}\right].
\end{split}
\end{equation}
Furthermore, let $P^1=[\beta^m_1]$. For $s=2,\cdots,m$, let the $s\times s$ binary matrix $\bs P^s$ be defined recursively as
\begin{equation}
\begin{split}\label{P}
\bs P^s=\left[\begin{array}{cccc}
  \bs P^{s-1} & \bs \alpha^s \\
  (\bs \alpha^s)^{\rm T} & \beta^m_s
\end{array}\right]
\end{split},
\end{equation}
where $[\beta^m_1,\beta_2^m,\cdots,\beta_s^m]^{\rm T}$ is the main diagonal elements of $\bs P^s$, and $\bs\alpha^s$ is a length $s-1$ column vector.

We have the following new result.
\begin{lem}\label{theorem1}
Given a length-$2^m$ RM sequence $\bs c^m$, its order $s$ and $s-1$ sub-sequences satisfy
\begin{equation}
\begin{split}\label{structure}
    \left\{\begin{array}{l}
      c^s_{2j}=v_j^{s-1}c_j^{s-1} \\
      c^s_{2j-1}=c_j^{s-1}
    \end{array}\right.,\quad j=1,\cdots,2^{s-1}, s=2,\cdots,m
\end{split}
\end{equation}
where
\begin{align}\label{v}
v_j^{s-1}&=(-1)^{b_s^m+\frac12\beta_s^m+(\bs \alpha^s)^{\rm T}\bs a_{j-1}^{s-1}}.
\end{align}
The vector $[v_1^{s-1},\cdots,v_{2^{s-1}}^{s-1}]^{\rm T}$ is a length-$2^{s-1}$ Walsh sequence with frequency $\bs\alpha^s$.

\end{lem}

\begin{IEEEproof}
Recall $a_j^s$ is the $s$-bit expression of $j$. For $1\le j\le 2^{s-1}$, the vector $\bs a_{2j-1}^s$ can be decomposed as
\begin{equation}
\begin{split}
    \bs a_{2j-1}^s=\left[\begin{array}{c}
               \bs a_{j-1}^{s-1} \\
               1
             \end{array}\right].
\end{split}
\end{equation}
Consequently
\begin{align}
    &2(\bs b^s)^{\rm T}\bs a_{2j-1}^s+(\bs a_{2j-1}^s)^{\rm T}\bs P^s\bs a_{2j-1}^s\notag\\
&= \left[\begin{array}{cc}
                                         \left(\bs a_{j-1}^{s-1}\right)^{\rm T} & 1
                                       \end{array}\right]
           \left[\begin{array}{cccc}
  \bs P^{s-1} & \bs \alpha^s \\
  (\bs \alpha^s)^{\rm T} & \beta^m_s
\end{array}\right]\left[\begin{array}{c}
               \bs a_{j-1}^{s-1} \\   1
             \end{array}\right]+2(\bs b^s)^{\rm T}\left[\begin{array}{c}
               \bs a_{j-1}^{s-1} \\   1
             \end{array}\right]\\\label{2j1}
&=2(\bs b^{s-1})^{\rm T}\bs a_{j-1}^{s-1}\!+2b_s^m\!+\beta_s^m+\!2(\bs \alpha^s)^{\rm T}\bs a_{j-1}^{s-1}\!+\left(\bs a_{j-1}^{s-1}\right)^{\rm T}\!\bs P^{s-1}\bs a_{j-1}^{s-1}.
\end{align}
Substituting \eqref{2j1} into \eqref{cm2} yields
\begin{align}
c^s_{2j}&=c_j^{s-1}i^{2b_s^m+\beta_s^m+2(\bs \alpha^s)^{\rm T}\bs a_{j-1}^{s-1}}\\
&=v_j^{s-1}c_j^{s-1}.
\end{align}

Likewise, the binary vector $\bs a_{2j-2}^s$ can be decomposed as
\begin{equation}
\begin{split}
    \bs a_{2j-2}^s=\left[\begin{array}{c}
               \bs a_{j-1}^{s-1} \\
               0
             \end{array}\right]
\end{split}.
\end{equation}
Then the exponent of $c^s_{2j-1}$ is expressed as
\begin{align}
    &2(\bs b^s)^{\rm T}\bs a_{2j-2}^s+(\bs a_{2j-2}^s)^{\rm T}\bs P^s\bs a_{2j-2}^s\notag\\
&=\left[\begin{array}{cc}
                                         \left(\bs a_{j-1}^{s-1}\right)^{\rm T} & 0
                                       \end{array}\right]
           \left[\begin{array}{cccc}
  \bs P^{s-1} & \bs \alpha^s \\
  (\bs \alpha^s)^{\rm T} & \beta_s^m
\end{array}\right]\left[\begin{array}{c}
               \bs a_{j-1}^{s-1} \\   0
             \end{array}\right]+ 2(\bs b^s)^{\rm T}\left[\begin{array}{c}
               \bs a_{j-1}^{s-1} \\  0
             \end{array}\right]\\\label{2j}
&=2(\bs b^{s-1})^{\rm T}\bs a_{j-1}^{s-1}+\left(\bs a_{j-1}^{s-1}\right)^{\rm T}\bs P^{s-1}\bs a_{j-1}^{s-1}.
\end{align}
Substituting \eqref{2j} into \eqref{cm2} yields
\begin{equation}
\begin{split}
    c^s_{2j-1}=c_j^{s-1}.
\end{split}
\end{equation}
\end{IEEEproof}

\section{Device Identification/Decoding and Channel Estimation}\label{useridentification}
In this section, we propose a novel RM detection algorithm for active device detection and channel estimation that leverages Theorem \ref{theorem1}.

Specifically, the matrix-vector pair of each message will be estimated recursively.
According to \eqref{b} and \eqref{P}, the matrix-vector pair $(\bs P^m,\bs b^m)$ is determined by $(\bs\alpha^{s},b_s^m,\beta_s^m), s=m,\cdots,2$ and $(b_{1}^m,\beta_{1}^m)$.
We will show that the algorithm first estimates the triple set $(\bs \alpha^{m},b_m^m,\beta_m^m)$, then $(\bs \alpha^{m-1},b_{m-1}^m,\beta_{m-1}^m)$, and finally the channel coefficient $h$ and $(b_{1}^m,\beta_{1}^m)$.

\subsection{Estimation of $(\bs \alpha^{m},b_{m}^m,\beta_{m}^m)$}
From \eqref{systemmodel} and \eqref{structure}, when $j=1,\cdots,2^{m-1}$, we have
\begin{align}
    y_{2j}^m&=\sqrt{\gamma}\sum_{k=1}^K h_k c_{k,2j}^m+z_{2j}^m\\\label{y2j0}
    &=\sqrt{\gamma}\sum_{k=1}^K h_k v_{k,j}^{m-1}c_{k,j}^{m-1}+z_{2j}^m,
\end{align}
and
\begin{align}
    y_{2j-1}^m&=\sqrt{\gamma}\sum_{k=1}^K h_k c_{k,2j-1}^m+z_{2j-1}^m\\\label{y2j1}
    &=\sqrt{\gamma}\sum_{k=1}^K h_k c_{k,j}^{m-1}+z_{2j-1}^m.
\end{align}

Since $c_{k,j}^{m-1}\in \{1,-1,i,-i\}$, we have $|h_kc_{k,j}^{m-1}|=|h_k|$. Define $\tilde y_j^{m-1}=y_{2j}^m(y_{2j-1}^m)^{*}$, \eqref{y2j0} and \eqref{y2j1} lead to
\begin{align}\label{ym1}
    \tilde y_j^{m-1}&=\gamma\sum_{k=1}^K |h_kc_{k,j}^{m-1}|^2 v_{k,j}^{m-1} +  \tilde {z}_{j}^{m-1}\\
    &=\gamma\sum_{k=1}^K |h_k|^2 v_{k,j}^{m-1} +  \tilde {z}_{j}^{m-1},
\end{align}
where
\begin{equation}
\begin{split}
    \tilde {z}_{j}^{m-1}&= \gamma\sum_{l=1}^K\sum_{k\ne l} h_kv_{k,j}^{m-1}c_{k,j}^{m-1}(h_lc_{l,j}^{m-1})^{*}\\
    &+\sqrt{\gamma}(z_{2j-1}^m)^{*}\sum_{k=1}^K h_k v_{k,j}^{m-1}c_{k,j}^{m-1}\\
    &+\sqrt{\gamma} z_{2j}^m\sum_{l=1}^K (h_lc_{l,j}^{m-1})^{*} +z_{2j}^m(z_{2j-1}^m)^{*}.
\end{split}
\end{equation}
The first term in the right-hand side of \eqref{ym1} is a linear combination of Walsh functions $v_{k,j}^{m-1}, k=1,2,\cdots,K$, with frequency $\bs\alpha^m_k$, which can be recovered by applying Walsh-Hadamard Transformation (WHT). The second term, $\tilde {z}_{j}^{m-1}$, is a linear combination of chirps, which can be considered to be distributed across all Walsh functions to equal degree, and therefore these cross-terms appear as a uniform noise floor.

Let the Hadamard matrix be $\bs W^m=\left[\bs w_1^m,\bs w_2^m,\cdots, \bs w_{2^m}^m\right]^{\rm T}$ and its $(l,j)$-th elements are $\bs W_{l,j}^m=(-1)^{(\bs a_{l-1}^m)^{\rm T}{\bs a_{j-1}^m}}$. Denote the WHT transformation as $\bs t^{m-1}=\bs W^{m-1}\tilde {\bs y}^{m-1}$, whose $l$-th entry can be written as
\begin{align}
    t_l^{m-1}&=(\bs w_l^{m-1})^{\rm T}\tilde {\bs y}^{m-1}\\
    &=\sum_{j=1}^{2^{m-1}}(-1)^{(\bs a_{l-1}^{m-1})^{\rm T}{\bs a_{j-1}^{m-1}}}\left(\gamma\sum_{k=1}^K |h_k|^2 v_{k,j}^{m-1} \!+\!  \tilde {z}_{j}^{m-1}\right)\\
    &=\gamma\sum_{j=1}^{2^{m-1}}(-1)^{\left(\bs a_{l-1}^{m-1}\right)^{\rm T}{\bs a_{j-1}^{m-1}}}\sum_{k=1}^K |h_k|^2 (-1)^{b_{k,m}^m+\frac12\beta_{k,m}^m+(\bs \alpha_k^m)^{\rm T}\bs a_{j-1}^{m-1}}\notag\\
    &+\sum_{j=1}^{2^{m-1}}(-1)^{(\bs a_{l-1}^{m-1})^{\rm T}{\bs a_{j-1}^{m-1}}}\tilde {z}_{j}^{m-1}\\
    &=\gamma\sum_{k=1}^K(-1)^{b_{k,m}^m+\frac12\beta_{k,m}^m} |h_k|^2\sum_{j=1}^{2^{m-1}}(-1)^{\left(\bs \alpha_k^{m}+\bs a_{l-1}^{m-1}\right)^{\rm T}{\bs a_{j-1}^{m-1}}}\notag\\
    &+\sum_{j=1}^{2^{m-1}}(-1)^{(\bs a_{l-1}^{m-1})^{\rm T}{\bs a_{j-1}^{m-1}}}\tilde {z}_{j}^{m-1}\label{hadaym1}.
\end{align}
Equation \eqref{hadaym1} indicates that, if we have $\bs \alpha_k^m=\bs a_{l-1}^{m-1}$, peaks will appear at frequency $\bs\alpha^m_k, k\in\{1,2,\cdots,K\}$, where the maximum value is $(-1)^{b_{k,m}^m+\frac12\beta_{k,m}^m}2^{m-1}\gamma |h_k|^2$.
On this basis, $ \bs{\hat\alpha}_k^m$ can be recovered by searching the largest absolute value of $\bs t^{m-1}$.
We further have
\begin{align}\label{bmbetam}
    (-1)^{b_{k,m}^m+\frac12\beta_{k,m}^m}=\left\{\begin{array}{rc}
                                        -i & \ {\rm if}\ (b_{k,m}^m,\beta_{k,m}^m)=(1,1), \\
                                        -1 & \ {\rm if}\ (b_{k,m}^m,\beta_{k,m}^m)=(1,0), \\
                                        i & \ {\rm if}\ (b_{k,m}^m,\beta_{k,m}^m)=(0,1), \\
                                        1 & \ {\rm if}\ (b_{k,m}^m,\beta_{k,m}^m)=(0,0).
                                      \end{array}\right.
\end{align}
Equation \eqref{bmbetam} indicates that $(b_{k,m}^m,\beta_{k,m}^m)$ can be estimated by the polarity of the largest value of $\bs t^{m-1}$.
For example, if the real part of the maximum value is positive and greater than the absolute value of the imaginary part, then we have $(b_{k,m}^m,\beta_{k,m}^m)=(0,0)$.

Further, $v_{k,j}^{m-1}$ is recovered through
\begin{equation}
\begin{split}\label{vkm1}
    {\hat v}_{k,j}^{m-1}=(-1)^{\hat b_{k,m}^m+\frac12\hat\beta_{k,m}^m+(\bs {\hat\alpha}_k^m)^{\rm T}\bs a_{j-1}^{m-1}}.
\end{split}
\end{equation}

\subsection{Estimation of $(\bs \alpha^{m-1},b_{m-1}^m,\beta_{m-1}^m)$}
After recovering $(\bs{\hat\alpha}_k^{m},b_{k,m}^m,\beta_{k,m}^m)$, we next estimate $(\bs\alpha_k^{m-1},b_{k,m-1}^m,\beta_{k,m-1}^m)$ in a similar way.
Define
\begin{equation}
\begin{split}\label{ym1next}
    y_j^{m-1}=\frac12\left({y_{2j-1}^m}+({\hat v}_{k,j}^{m-1})^{*} y_{2j}^m\right).
\end{split}
\end{equation}
Under the assumption that ${\hat v}_k^{m-1}$ is correctly estimated, according to \eqref{y2j0} and \eqref{y2j1}, $y_j^{m-1}$ is further expressed as
\begin{equation}
\begin{split}\label{yjm1}
    y_{j}^{m-1}\!&=\sqrt{\gamma}h_k c_{k,j}^{m-1}\!+\!\frac{\sqrt{\gamma}}2\sum_{l\ne k}\! h_l c_{l,j}^{m-1}\!\left(\! 1+(\hat v_{k,j}^{m-1})^{*}v_{l,j}^{m-1}\right) +z_j^{m-1},
\end{split}
\end{equation}
where
\begin{equation}
\begin{split}
    z_j^{m-1}=\frac12\left((\hat v_{k,j}^{m-1})^{*}z_{2j}^m+z_{2j-1}^m\right)\sim {\cal CN}\left(0,\frac{1}{2}\right),\\
\end{split}
\end{equation}
i.e., the variance of the channel noise is reduced by half.
Since $\hat {\bs v}_{k}^{m-1}$ and $\bs v_{l}^{m-1}$ are two different Walsh functions, $\hat v_{k,j}^{m-1}v_{l,j}^{m-1}$ take vales in $\{1,-1,i,-i\}$ with equal probability, in other words, $P\left(1+(\hat {v}_{k,j}^{m-1})^{*}v_{l,j}^{m-1}=0\right)=\frac14$.
This indicates that the interference from other devices is reduced by a quarter.

For simplicity, let $u_{l,j}^{m-1}=\frac12\left(\! 1+(\hat v_{k,j}^{m-1})^{*}v_{l,j}^{m-1}\right)$, and we have $u_{k,j}^{m-1}=1$. Given this, \eqref{yjm1} leads to
\begin{align}
    y_{j}^{m-1}\!&=\sqrt{\gamma}h_k c_{k,j}^{m-1}\!+\!{\sqrt{\gamma}}\sum_{l\ne k}\! h_l c_{l,j}^{m-1}u_{l,j}^{m-1} +z_j^{m-1}\\\label{yjmu}
    &={\sqrt{\gamma}}\sum_{l=1}^K\! h_l c_{l,j}^{m-1}u_{l,j}^{m-1} +z_j^{m-1}.
\end{align}
Compared with \eqref{systemmodel}, the device interference from other devices is reduced by a quarter and the variance of the noise is reduced by half.

When $1\le j\le 2^{m-2}$, applying Theorem.\ref{theorem1} on \eqref{yjmu} leads to
\begin{align}
    y_{2j}^{m-1}&=  \sqrt{\gamma}{\mathop \sum\limits_{l=1}^K} h_l c_{l,2j}^{m-1}u_{l,2j}^{m-1}+z_{2j}^{m-1}\\
    &= \sqrt{\gamma}\sum_{l=1}^K h_l v_{l,j}^{m-2}c_{l,j}^{m-2}u_{l,2j}^{m-1}+z_{2j}^{m-1}
\end{align}
and
\begin{align}
    y_{2j-1}^{m-1}&=\sqrt{\gamma}\sum_{l=1}^K h_l c_{l,{2j-1}}^{m-1}u_{l,2j}^{m-1}+z_{2j-1}^{m-1}\\
    &= \sqrt{\gamma}\sum_{l=1}^K h_l c_{l,j}^{m-2}u_{l,2j}^{m-1}+z_{2j-1}^{m-1}.
\end{align}
Let $\tilde{y}_j^{m-2}=y_{2j}^{m-1}(y_{2j-1}^{m-1})^{*}$, we have
\begin{equation}
\begin{split}\label{yjm21}
    \tilde{y}_j^{m-2}&=\gamma\sum_{l=1}^K h_l v_{l,j}^{m-2}c_{l,j}^{m-2}u_{l,2j}^{m-1}\sum_{q=1}^K \left(h_q c_{q,j}^{m-2}u_{q,2j}^{m-1}\right)^{*}+\tilde z_{j}^{m-2},\\
\end{split}
\end{equation}
where $v_{l,j}^{m-2}=(-1)^{b_{l,m-1}^m+\frac12\beta_{l,m-1}^m+(\bs \alpha_{l}^{m-1})^{\rm T}\bs a_{j-1}^{m-2}}$ and
\begin{equation}
\begin{split}
    \tilde z_{j}^{m-2}&=\sqrt{\gamma}(z_{2j-1}^{m-1})^{*}\sum_{l=1}^K h_l v_{l,j}^{m-2}c_{l,j}^{m-2}u_{l,2j}^{m-1}\\
    &  +\sqrt{\gamma}z_{2j}^{m-1}\sum_{q=1}^K \left(h_q c_{q,j}^{m-2}u_{l,2j}^{m-1}\right)^{*}+z_{2j}^{m-1}(z_{2j-1}^{m-1})^{*}.\\
\end{split}
\end{equation}
Similar to \eqref{ym1}, applying WHT on \eqref{yjm21}, $\bs {\hat\alpha}_k^{m-1}$ can be recovered by searching the maximum value of the result. Comparing \eqref{ym1} and \eqref{yjm21}, we know that $\bs {\hat\alpha}_k^{m-1}$ is more likely to be correctly estimated than $\bs {\hat\alpha}_k^{m}$ because the interference terms are reduced by a quarter and the variance of channel noise is reduced by half.

Similar to \eqref{bmbetam}, $(b_{k,m-1}^m,\beta_{k,m-1}^m)$ can be recovered by the polarity of the maximum value.

\subsection{Estimation of Channel Coefficient $(b_{1}^m,\beta_{1}^m,h)$}
We continue these process until all the estimates $\left(\bs {\hat\alpha}_k^s, {\hat b}_{k,s}^m, {\hat\beta}_{k,s}^m\right), s\in\{2,\cdots,m\}$ are obtained.
According to \eqref{yjm1}, the received sequence in the last layer can be written as
\begin{align}
    y_{j}^{1}&=\sqrt{\gamma} h_k c_{k,j}^{1}+A+z_j^{1}, \qquad j=1,2,
\end{align}
where the term $A$ consists of all {\em interferences} from other devices which are all second order RM sequences and $z_j^{1}\sim{\cal}(0,\frac1{2^{m-1}})$.
Accordingly, we have
\begin{align}\label{y11}
    y_1^1&=\sqrt{\gamma}h_k c_{k,1}^1+A+z_{1}^1\\
    &=\sqrt{\gamma}h_k +A+z_{1}^1
\end{align}
and
\begin{align}\label{y21}
    y_2^1&=\sqrt{\gamma}h_k c_{k,2}^1+A+z_{2}^1\\
    &=\sqrt{\gamma}h_k(-1)^{b_{k,1}^m+\frac12\beta_{k,1}^m}+A+z_{2}^1.
\end{align}
Similar to previous processing, define $\tilde{y}_1^0=(y_1^1)^{*}y_2^1$, we have
\begin{align}\label{y10}
    \tilde{y}_1^0={\gamma}|h_k|^2(-1)^{b_{k,1}^m+\frac12\beta_{k,1}^m}+\tilde z_1^0,
\end{align}
where
\begin{align}
    \tilde{z}_1^0=\left[\sqrt{\gamma}h_k(-1)^{b_{k,1}^m+\frac12\beta_{k,1}^m}+A+z_{2}^1\right]\left[A^{*}+(z_1^1)^{*}\right]+\sqrt{\gamma}h_k^{*}(A+z_2^1).
\end{align}
According to \eqref{y10}, $(b_{k,1}^m,\beta_{k,1}^m)$ can be estimated by the polarity of $\tilde{y}_1^0$.

Further the channel coefficient $h_k$ can be estimated as
\begin{equation}
\begin{split}\label{hk}
    \sqrt{\gamma}\hat h_k=\frac1{2}\left({y_1^1+\left((-1)^{\hat b_{k,1}^m+\frac12\hat\beta_{k,1}^m}\right)^{*}y_2^1}\right).
\end{split}
\end{equation}
Note that we can infer $\hat h_k$ from \eqref{hk} if the AP knows the value of $\gamma$, but we only need to know the product of the two.

\subsection{RM Detection Algorithm}
So far, the matrix-vector pair $\left(\hat{\bs P}_k^m,\hat{\bs b}_k^m\right)$ and the corresponding channel coefficient $\hat h_k$ for message $k$ has been completely estimated. In addition, $\hat{\bs c}_k^m$ can be obtained through \eqref{cm2}. Given this, the remaining messages can be estimated iteratively by updating the received signal $\bs y^m$ by removing the interference of device $k$.

The detailed algorithm is summarized in Algorithm 1, where the maximum number of detected devices is limited to $K_{\rm max}$.
The choice of $K_{\rm max}$ depends on the expected success rate, miss rate, and false alarm rate. On one hand, a small value of $K_{\rm max}$ will lead to an unsatisfactory success rate and miss rate. On the other hand, if the value of $ K_{\rm max} $ is too large, it may lead to a higher false alarm rate and computational complexity.
Hence, the choice of $K_{\rm max}$ is determined experimentally by the tradeoff between the success rate, miss rate, and the false alarm rate, as well as the practical complexity constraint.

From the above analysis, on condition that $\bs\alpha^m$ is correctly estimated, $\bs\alpha^{s}, s=m-1,m-2,\cdots,2$, are more likely to be correctly estimated than $\bs\alpha^{m}$ because both interference and channel noise are reduced. However, if $\bs\alpha^m$ is wrongly estimated, the desired message will also be reduced by half, which will likely cause an error propagation problem that leads to the wrong estimation of $\bs\alpha^{s}, s=m-1,m-2,\cdots,2$.

\begin{center}
\begin{tabular}{l l}
\toprule
\multicolumn{1}{l}{{\bf Algorithm 1}: RM detection algorithm.}\\
\midrule
{\bf Input}: the received signal $\bs y^m$, the maximum number of active devices $K_{\rm max}$.\\
$k=0$.\\
{\bf while} $k<K_{\rm max}$ and $\|\bs y^m\|^2>\varepsilon $\footnotemark[1] {\bf do}\\
\quad $k\leftarrow k+1$.\\
\quad {\bf for }$s=m,m-1,\cdots,2$ {\bf do}\\
\qquad Split $\bs y^m$ into two partial sequences according to \eqref{y2j0} and \eqref{y2j1}.\\
\qquad Perform the element-wise conjugate multiplication according to \eqref{ym1}.\\
\qquad Perform WHT and let $\hat{\bs \alpha}^s$ be the binary index of the largest component.\\
\qquad Recover $(\hat b_{s}^m,\hat \beta_{s}^m)$ according to \eqref{bmbetam}.\\
\qquad Recover $\hat{\bs v}^{s-1}$ according to \eqref{vkm1}.\\
\qquad Calculate $\bs y^{m-1}$ according to \eqref{ym1next}.\\
\quad {\bf end for}\\
\quad Recover $(\hat b_{1}^m,\hat \beta_{1}^m)$ according to \eqref{bmbetam} and \eqref{y10}.\\
\quad Add $(\hat{\bs P}^m, \hat{\bs b}^m)$ to the decoded set.\\
\quad Calculate the codeword $\hat{\bs c}^m$ according to $(\hat{\bs P}^m, \hat{\bs b}^m)$ and estimate $\hat h$ according to \eqref{hk}.\\
\quad $\bs y^m\leftarrow \bs y^m-\sqrt{\gamma}\hat h\bs c^m$.\\
{\bf end while}\\
{\bf Output}: $\left(\hat{\bs P}_l^m, \hat{\bs b}_l^m\right)$, $\hat{\bs c}_l^m$ and $\hat h_l$ for $l=1,2,\cdots, k$.\\
\bottomrule
\end{tabular}
\footnotetext[1]{Various criteria are possible here. We found $\varepsilon=(2^{m/2}+2)^2$ works well when there are only in-cell devices transmit, and $\varepsilon=2\sigma^2+2^{m+1}$ works well when there are transmission from out-of-the-cell devices. We should emphasize that it's very much an open question as to how to choose this parameter optimally. Our suggestion would be to try to optimize the value empirically for the regime of interest.}
\end{center}

~\\

To avoid error propagation, we propose a list detection algorithm inspired by \cite{Hanzo}. Let $\bs L=[L_m, L_{m-1}, \cdots, L_{m-|L|+1}]$ where $L_s, m-|L|+1\le s\le m$, denotes the number of decoding paths in layer $s$. On this basis, when estimating $(\bs\alpha^s,b^m_s,\beta^m_s)$, instead of choosing the largest component, we maintain the largest $L_s$ component as candidates. Thus the total number of decoding paths is $\prod_{s=m}^{m-|L|+1}L_s$.
And finally, the path having the minimum residual energy is chosen as the optimal path where the residual energy of path $l\in\{1,\prod_{s=m}^{m-|L|+1}L_s\}$ is defined as
\begin{equation}
\begin{split}\label{residualenergy}
    R_l=\left\|\bs y^m-\sqrt{\gamma}\hat h_l\hat{\bs c}_l^m\right\|^2.
\end{split}
\end{equation}

\begin{figure}[H]
  \begin{center}
  \includegraphics[width=2.8in]{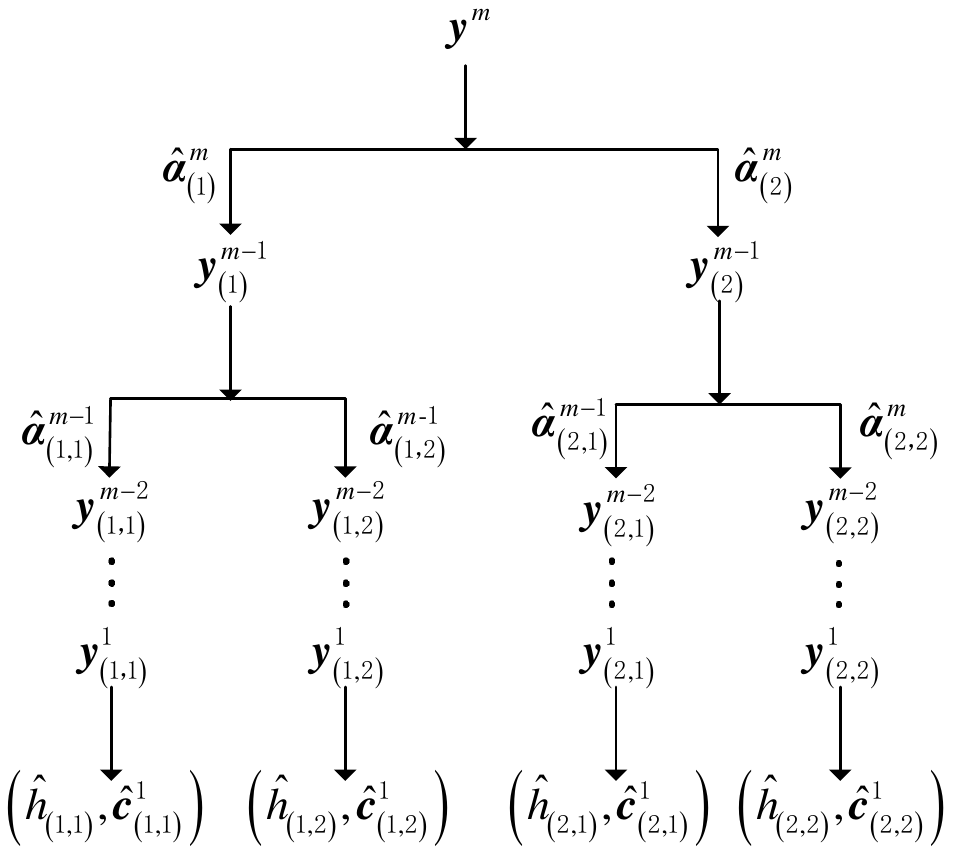}
  \end{center}
  \caption{An example of list decoding with $\bs L=[2,2]$.}\label{fig1}
\end{figure}

An example of $\bs L=[2,2]$ list decoding algorithm is illustrated in Fig.\ref{fig1}.

\section{Enhanced RM Decoding Algorithm with Slotting and Message Passing}\label{algorithm2}
Inspired by \cite{Robert}, we adopt the idea of \emph{slotting} and \emph{message passing} to enhance the performance of Algorithm 1.
In general, slotting trades bandwidth for better performance.
\subsection{Slotting}
By slotting, we mean that a codeword of length $n=2^{m}$ is divided into $2^p$ slots, and each message of length $2^{m-p}=2^q$ is transmitted in one or more of the $2^p$ slots.
Accordingly, each message is assign a sparse codeword which is zero except for the slots to which the message is sent.
Then we estimate the message within each slot, and combine the results.

In practice, we send each message in two out of the $2^p$ slots.
We append $p$ bits to encode a primary slot location.
From the $\frac12q(q+3)$ bits used to encode $(\bs P^q, \bs b^q)$, we use an arbitrary subset of size $p$ to encode a \emph{translate}, which gives the secondary slot location when it is added to the primary slot location.
To distinguish the primary and secondary slots, we fix a signal \emph{check bit} in the $\bs P^q$ matrix to be 0 for the primary slot and 1 for the secondary slot. Thus deducing 1 bit from the total number of bits transmitted.
Thus, the number of bits that each message transmitted is
\begin{equation}
\begin{split}\label{B}
    B=\frac12q(q+3)+p-1.
\end{split}
\end{equation}
Given $m$ (which determines the codelength), the larger $p$ is, the more significant the reduction in the number of information bits, the better the error performance.

\subsection{Message Passing}

By transmitting each message in two slots, the decoded messages in one slot are expected to be propagated to the other slots.
The decoding algorithm cycles through all the slots. Whenever a $(\hat{\bs P}^q, \hat {\bs b}^q, {\hat h})$ set is found, the algorithm first checks whether the corresponding message has been recorded. If not recorded, the check bit in the $\hat{\bs P}^q$ matrix reveals whether the current slot is primary or secondary. If the current slot is the primary slot (secondary slot), the location of the secondary slot (primary slot) is determined by the current slot plus the translate that implied in the $(\hat{\bs P}^q, \hat{\bs b}^q)$ pair. Then after flipping the check bit in the $\bs {\hat P}^q$ matrix, the decoded $(\bs {\hat P}^q, \bs {\hat b}^q, {\hat h})$ set is propagated to the other slot to which the message was sent.

\subsection{RM Detection Algorithm}

Upon receiving the messages, decoding the slots one by one, i.e., decoding the received signal in slot 1 first, and then slot 2, and so on.
The detailed algorithm is summarized as in Algorithm 2.

~\\
\noindent
\begin{center}
\begin{tabular}{l l}
\toprule
\multicolumn{1}{l}{{\bf Algorithm 2}: Enhanced RM detection algorithm.}\\
\midrule
{\bf Input}: the received signal $\bs Y\!\!=\![\bs y^q_1,\cdots,\bs y^q_{2^p}]$, the average number of devices $\overline K_{\rm max}$ in each slot.\\
$\bs P=[\ ]$, $\bs b=[\ ]$, $slot=[\ ], h=[\ ]$, $\bs c=[\ ]$, $l=0$.\\
{\bf for} $i=1:2^p$ {\bf do}\\
\quad $k\leftarrow0$.\\
\quad {\bf for }$j=1:l$ {\bf do}\\
\quad\quad {\bf if} $slot[j]=i$ {\bf do}\\
\quad\quad\quad $\bs y^q_i\leftarrow \bs y^q_i-\bs c[:,j]\times h[j]$.\\
\quad\quad\quad $k\leftarrow k+1$.\\
\quad\quad {\bf end if}\\
\quad {\bf end for}\\
\quad Input $\overline K_{\rm max}-k$ as the updated maximum number of devices and $\bs y_i^q$ to Algorithm 1,\\
\quad and record the outputs. Let $k_1$ be the number of recovered messages.\\
\quad {\bf for} $j=1:k_1$\\
\quad\quad {\bf if } $(\hat{\bs P}_j^q, \hat{\bs b}_j^q)$ are not recorded in $(\bs P, \bs b)$ {\bf do}\\
\quad\quad\quad $l\leftarrow l+1$.\\
\quad\quad\quad $\bs P[:,:,l]\leftarrow \hat{\bs P}_j^q$.\\
\quad\quad\quad $\bs b[:,l]\leftarrow \hat{\bs b}_j^q$.\\
\quad\quad\quad $h[l]\leftarrow \hat{h}_j^q$\\
\quad\quad\quad $\bs c[:,l]\leftarrow \hat{\bs c}_j^q$.\\
\quad\quad\quad Calculate the translate according to $\left(\hat{\bs P}_j^q, \hat{\bs b}_j^q\right)$ and update $slot[l]$.\\
\quad\quad {\bf end if }\\
\quad {\bf end for}\\

{\bf end for}\\
{\bf Output}: $\bs P, \bs b, h, \bs c$.\\
\bottomrule
\end{tabular}
\end{center}
~\\
\noindent

\section{Computational Complexity Analysis}\label{CCA}
For active device detection and channel estimation, Algorithm 1 iterates at most $K_{\rm max}$ times to obtain $K_{\rm max}$ messages. The most expensive operations in each iteration come from two parts: the calculation of WHT and the generation of the RM code through \eqref{cm2}.
We consider the number of multiplication operations required to run the algorithm, where the number of addition operations is similar to it.
In each iteration, the complexity of obtaining $\hat{\bs \alpha}^s$ using fast WHT is ${\cal O}(s2^s), s=m,m-1\cdots,2$. And thus the number of multiplication operations required for obtaining $\hat{\bs P}^m$ in each iteration is on the order of $\sum_{s=2}^m{\cal O}(s2^s)={\cal O}((2m-2)2^m)$.
Thus, the complexity of Algorithm 1 for calculating WHT is ${\cal O}(K_{\rm max}(2m-2)2^m)$.
Similarly, in each iteration, the complexity of generating the Reed-Muller code when a matrix-vector pair $\left(\hat{\bs P}^m, \hat{\bs b}^m\right)$ is found is ${\cal O}((m^2+m)2^m)$. Therefore, the complexity of Algorithm 1 for generating the Reed-Muller code is ${\cal O}(K_{\rm max}(m^2+m)2^m)$.
In summary, the worst-case complexity of Algorithm 1 is on the order of ${\cal O}(K_{\rm max}(m^2+3m-2)2^m)$.

To enhance the performance of Algorithm 1, we adopt the idea of slotting and message passing and proposed an enhanced algorithm summarized in Algorithm 2, as well as a call to Algorithm 1.
Namely, a length-$2^m$ codeword is divided into $2^p$ slots, where the length of the sequence in each slot is $2^{m-p}=2^q$.
Since the number of slots is $2^p$, the complexity of algorithm 2 is ${\cal O}(2^p\overline K_{\rm max}(q^2+3q-2)2^q)$.
In practice, the maximum number of the neighbors of the AP in each slot in Algorithm 2 is set to $\overline K_{\rm max}=\left\lceil\frac{3K^{*}}{2^{p-1}}\right\rceil$, where $\lceil\cdot\rceil$ stands for rounding up.
Accordingly, the complexity of Algorithm 2 is ${\cal O}(6K^{*}(q^2+3q-2)2^q)$.

For comparison, we summarize the computational complexity of the algorithm proposed in \cite{Hanzo} as follows.
For the sake of explanation, we denote the algorithm given in \cite{Hanzo} as the ``\emph{list RM\_LLD}'' algorithm.
The computational complexity of the list RM\_LLD algorithm also come from two parts: the calculation of WHT and the generation of the RM code through \eqref{cm2}. In each iteration, the complexity of fast WHT is ${\cal O}((2m-2)2^m)$, and there are totally $K_{\rm max}n_{\rm max}$ iterations in Algorithm 2, where $K_{\rm max}$ denotes the maximum number of active devices and $n_{\rm max}$ is the maximum number of iterations.
Thus, the complexity of applying fast WHT is ${\cal O}(K_{\rm max}n_{\rm max}(2m-2)2^m)$.
Likewise, the complexity of generating the RM code is ${\cal O}(K_{\rm max}n_{\rm max}(m^2+m)2^m)$.
Accordingly, the complexity of the list RM\_LLD algorithm is ${\cal O}(K_{\rm max}n_{\rm max}(m^2+3m-2)2^m)$.
In \cite{Hanzo}, the maximum number of iterations is set to be $n_{\rm max}=5$.

The comparison of the computational complexity of Algorithm 1, Algorithm 2, and the list RM\_LLD algorithm are summarized in Table \ref{table1}.

\begin{table}[htbp]
\caption{Comparison of the number of multiplication operations required.}
\centering
\renewcommand\arraystretch{1.2}
\begin{tabular}{l|l}
  \hline
  \label{table1}
    Algorithm 1 & ${\cal O}(K_{\rm max}(m^2+3m-2)2^m)$\\
    \hline
    Algorithm 2 & ${\cal O}(6K^{*}(q^2+3q-2)2^q)$\\
    \hline
    List RM\_LLD algorithm &${\cal O}(5K_{\rm max}(m^2+3m-2)2^m)$ \\
  \hline
\end{tabular}
\end{table}

Algorithm 1 and the algorithm in \cite{Hanzo} can be considered as a special case of Algorithm 2 with only one slot, i.e., $p=0, q=m$, where $K_{\rm max}$ denotes the maximum number of devices of the slot.
From Table \ref{table1}, it is observed that the complexity of the list RM\_LLD algorithm is 5 times larger than Algorithm 1 because it cycles to detect $K_{\rm max}$ active device for 5 times to get better performance.
What stands out in Table \ref{table1} is that the complexity of Algorithm 2 is more than $2^p$ times lower than that of Algorithm 1 and the algorithm in \cite{Hanzo}.


\section{Numerical Results}\label{numericlresult}
\subsection{Definition of Error Metrics}
We first define the false alarm rate, the miss rate, the success rate, and the channel estimation error. These are our main performance metrics.

\begin{figure}[H]
  \begin{center}
  \includegraphics[width=3in]{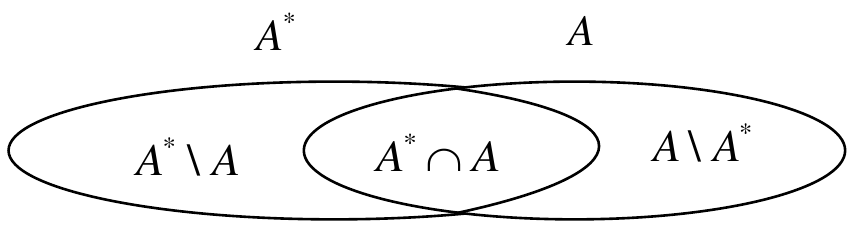}
  \end{center}
  \caption{The set relationship.}\label{setrelation}
\end{figure}

Denote $A^{*}\subset\{1,2,\cdots,2^B\}$ as the index of the messages transmitted by in-cell devices of the AP. We have $|A^{*}|=K^{*}$. And let $A\subset\{1,2,\cdots,2^B\}$ denotes the index of the output messages of the algorithm.
The set relationship is depicted in Fig. \ref{setrelation}.

The output of our algorithm is divided into two phases. In the first phase, the AP does not know the number of its in-cell devices $K^{*}$. In this case, the algorithm output all the detected messages in $A$.
Accordingly, we define the false alarm rate in this phase as
\begin{equation}
\begin{split}\label{DefiniFAR}
    \frac{|A\backslash A^{*}|}{|A|},
\end{split}
\end{equation}
and the miss rate as
\begin{equation}
\begin{split}\label{DefiniMAR}
    \frac{|A^{*}\backslash A|}{|A^{*}|}.
\end{split}
\end{equation}
The false alarm rate and miss rate reflect the performance of the algorithm when the number of devices are not known in the AP.

In the second phase, we assume that the AP knows the exact number of in-cell devices $K^{*}$.
If $|A|>K^{*}$, the algorithm chooses to output $K^{*}$ elements from the set $A$, with which the recovered channel gain of these $K^{*}$ devices is the strongest. Without loss of generality, denote the $K^{*}$ output elements as $A_{1:K^{*}}$; otherwise, it outputs $A$ directly.

We define the average per-user probability of successful decoding following \cite{Polyanskiy,Robert} assuming $K^{*}$ is known to gauge the performance of the algorithm.
Given $k\in A^{*}$, denote $E_k$ as an error event if $k\notin A_{1:K^{*}}$.
In this case, we define the success rate in the second phase as
\begin{equation}
\begin{split}\label{sdp}
    1-\frac1{K^{*}}\sum\limits_{k\in A^{*}} P(E_k).
\end{split}
\end{equation}
The definition of success rate assumes that the algorithm knows the number of in-cell devices of the AP $K^{*}$, which may actually be unrealistic.
In addition, in the numerical results section in \cite{Hanzo}, the authors assume that the number of in-cell devices of AP is known, i.e., the maximum number of detected devices $K_{\rm max}$ is set to equal to $K^{*}$. In that case, the false alarm rate and miss rate of the algorithm in \cite{Hanzo} are equal according to the definition in \eqref{DefiniFAR} and \eqref{DefiniMAR}, the success rate plus either false alarm rate or miss rate equals to 1.

We define the channel estimation error in the second phase as follows.
On condition that message $k$ is successfully decoded, i.e., given $k\in A^{*}$, we have $k\in A_{1:K^{*}}$, if the error between $h_k$ and its estimation $\hat h_k$ is greater than a given threshold $\theta_{h_k}$, then $h_i$ is considered to be wrongly estimated.
In this case, channel estimation error is defined as
\begin{equation}
\begin{split}\label{cee}
    P\left(\left|h_k-\hat h_k\right|>\theta_{h_k}\right).
\end{split}
\end{equation}
Note that the quality of channel estimation is usually measured by the mean square error \cite{Hanzo}, but in practice, if the estimation error is less than a certain value, the reception quality has almost no impact. Therefore, we use the channel estimation error defined in \eqref{cee} in this paper.

\subsection{Simulation Parameter Setting}
The performance comparison indicators are the false alarm rate, success rate, miss rate, and channel estimation error as defined above.
The number of list decoding paths for Algorithm 1, Algorithm 2, and the list RM\_LLD algorithm are all set to $L=4$.
The maximum number of detected devices in Algorithm 1 and the list RM\_LLD algorithm are set to $K_{\rm max}=K^{*}$. In other words, Algorithm 1 and the list RM\_LLD algorithm need to know the exact number of active devices when decoding the messages, which may not be practically realistic. In contrast, Algorithm 2 does not to know the number of active devices.

The detailed simulation parameters are listed in Table \ref{tablesimulationparameter}.

\begin{table}[htbp]
\caption{Simulation Parameters.}
\centering
\begin{tabular}{l|c}
  \hline
  \label{tablesimulationparameter}
  Parameters & Value \\
  \hline
  Channel gain threshold $\theta$ & $10^{-6}$ \\
  Path-loss exponent $\bs \alpha$ & $4$ \\
  Channel gain error estimation threshold $\theta_{h_k}$ for device $k$ & $0.3|h_k|$ \\
  Transmit SNR of each device $\gamma$ & 60 dB \\
  Number of list decoding paths in all algorithms $L$ & $4$ \\
  Average number of detected devices in each slot $\overline K_{\rm max}$ & $\left\lceil\frac{3K^{*}}{2^{p-1}}\right\rceil$ \\
  Maximum number of detected devices $K_{\rm max}$ & $K^{*}$ \\
  $n_{\rm max}$ in the list RM\_LLD algorithm & $5$ \\
  \hline
\end{tabular}
\end{table}

\subsection{Performance Comparison without Out-of-the-cell Devices}
In \cite{Hanzo}, the author assumes that there are no transmissions from out-of-the-cell devices, that is, all active devices are in-cell devices of AP.
To compare with the result given in \cite{Hanzo}, we assume $K=K^{*}$ in this subsection.

\begin{figure*}[htbp]
\centering

\subfigure[]{
\begin{minipage}[t]{0.45\linewidth}
\centering
\includegraphics[width=2.2in]{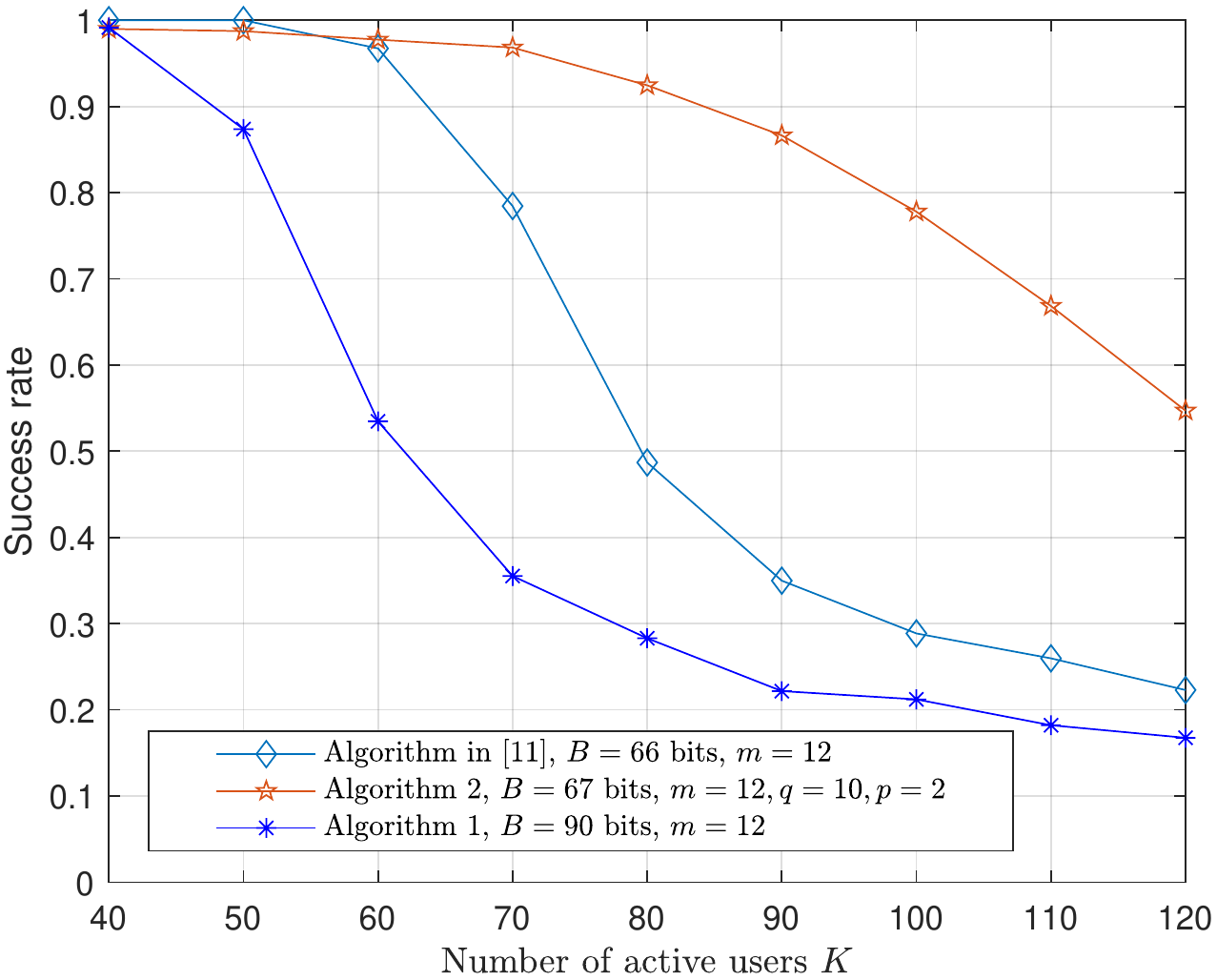}
\end{minipage}%
\label{SDP}
}%
\subfigure[]{
\begin{minipage}[t]{0.45\linewidth}
\centering
\includegraphics[width=2.2in]{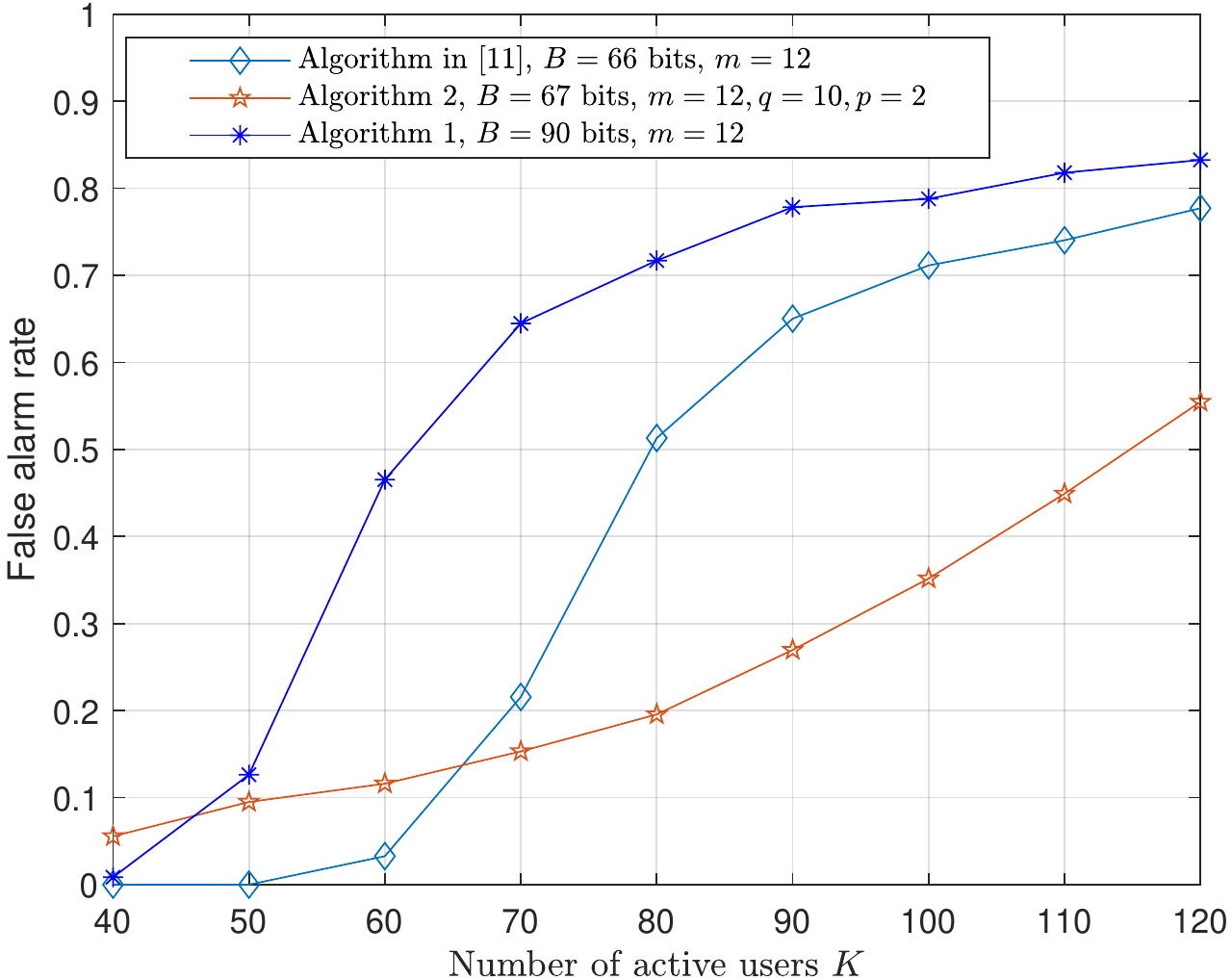}
\end{minipage}%
\label{FAR}
}%

\subfigure[]{
\begin{minipage}[t]{0.45\linewidth}
\centering
\includegraphics[width=2.2in]{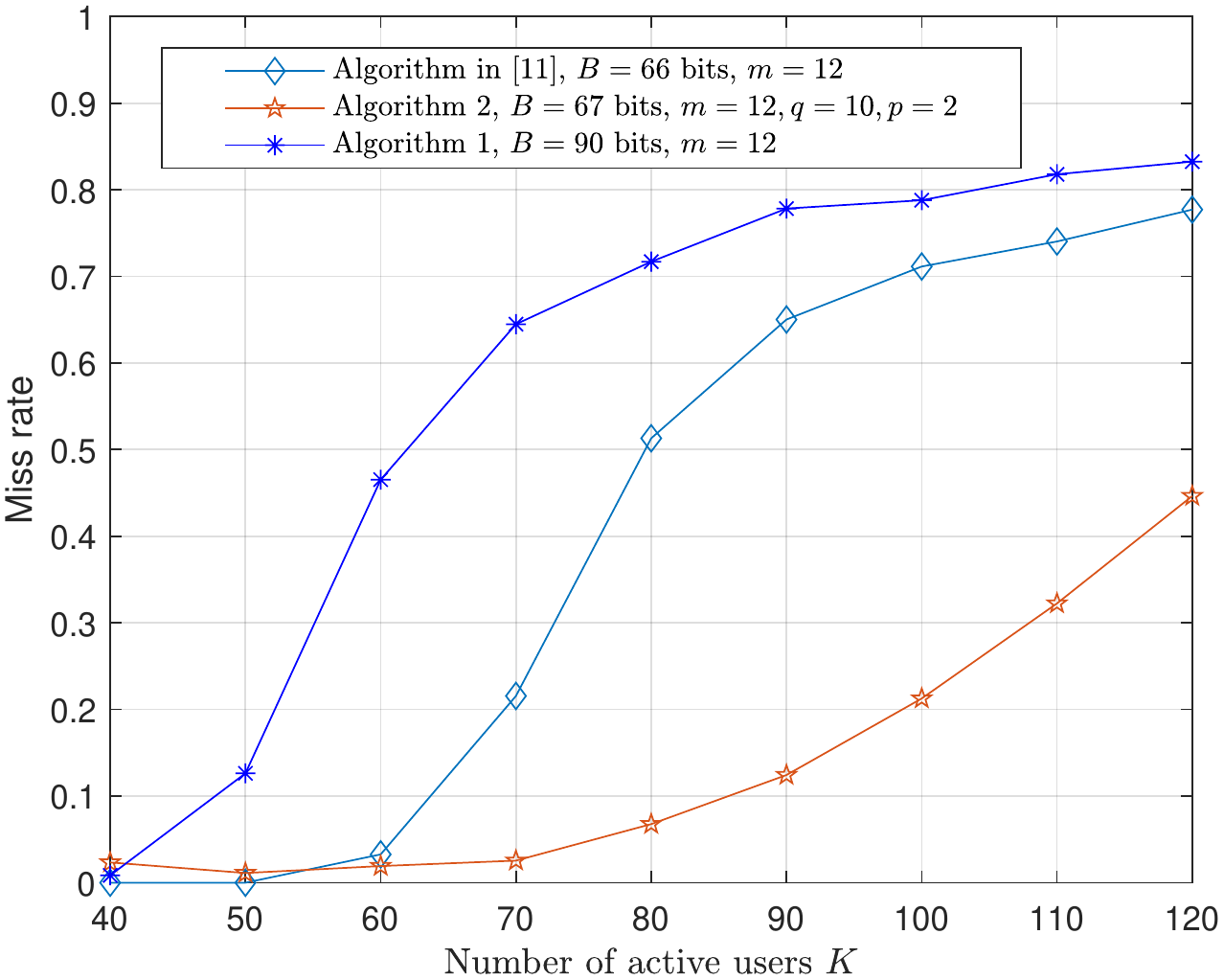}
\end{minipage}
\label{MAR}
}%
\subfigure[]{
\begin{minipage}[t]{0.45\linewidth}
\centering
\includegraphics[width=2.2in]{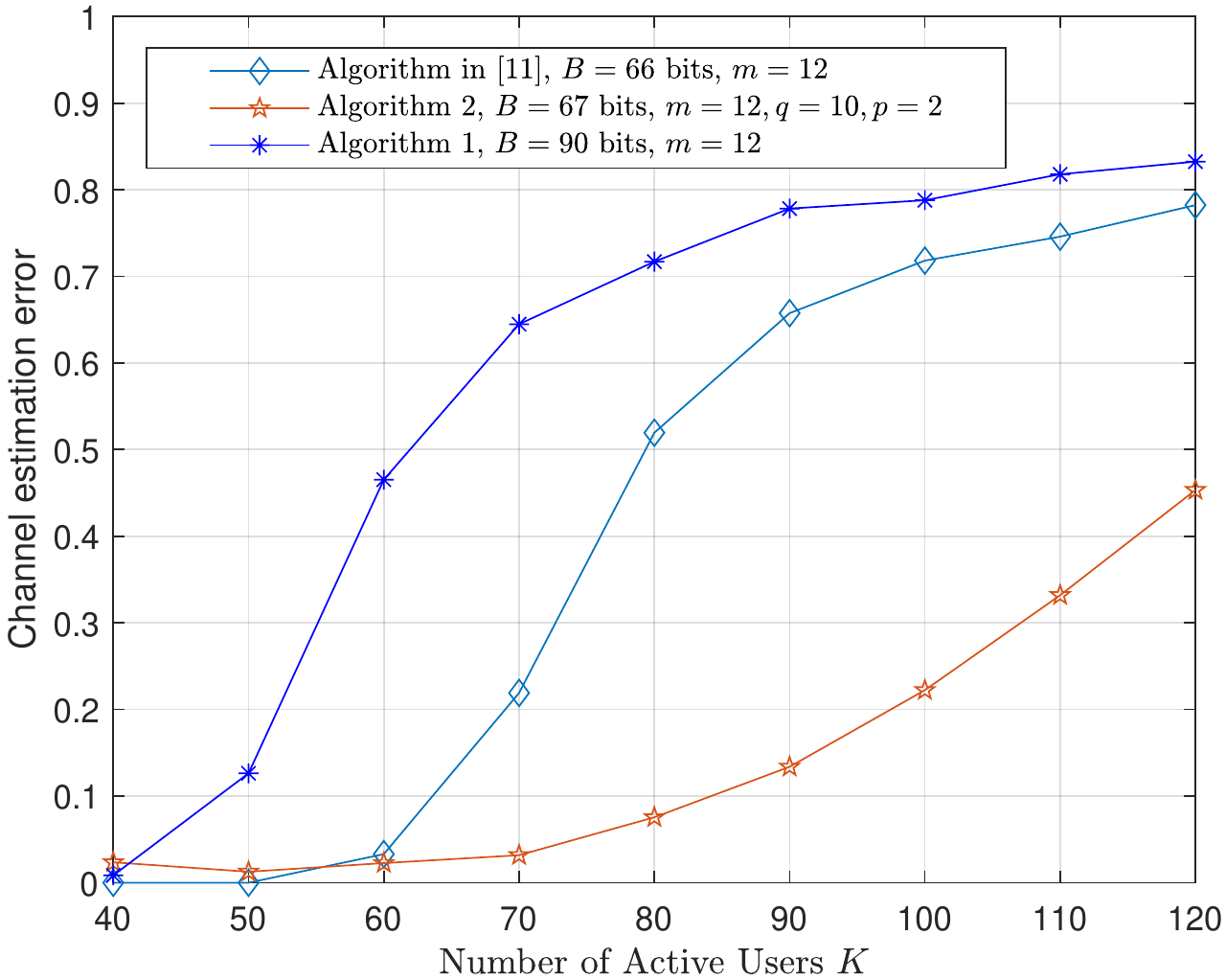}
\end{minipage}
\label{CEE}
}%

\centering
\caption{Performance comparison between different RM decoding algorithms. (a) The success rate versus $K$, (b) the false alarm rate versus $K$, (c) the miss rate versus $K$, and (d) the channel estimation error versus $K$.}
\label{figureallneighbor}
\end{figure*}

Fig. \ref{figureallneighbor} compares the performance of Algorithm 2 and the list RM\_LLD algorithm given in \cite{Hanzo}. We fix $m=12$ so that the codelength is $2^{12}=4,096$. To keep the transmit bits of Algorithm 2 and the list RM\_LLD algorithm comparable, we set $p=2, q=m-2$ in Algorithm 2. Since the number of slots is small, we do not use message passing in this case, i.e., each messages randomly chooses one of the 4 slots for transmitting. In this case, we don't need to fix the check bit in the $\bs P$ matrix to indicate whether the current slot is the primary or secondary slot. Thus increasing one information bit in \eqref{B}.
Accordingly, the number of transmit bits of Algorithm 2 is $B=\frac12q(q+3)+p=67$ bits, and the number of transmit bits of the list RM\_LLD algorithm in \cite{Hanzo} is $B=\frac12m(m-1)=66$ bits.

Fig. \ref{SDP} depicts the success rates of Algorithm 1, Algorithm 2, and list RM\_LLD algorithm versus the number of active devices.
It can be observed that when the number of active devices is greater than 60, the performance of Algorithm 1 and the list RM\_LLD algorithm degrades rapidly.
In contrast, the performance of Algorithm 2 degrades gracefully with the number of active devices.
In addition, Algorithm 2 performs much better than Algorithm 1 and the list RM\_LLD algorithm when $K$ is larger than 60.
Meanwhile, the list RM\_LLD algorithm outperforms Algorithm 1 over all ranges.
This is because the list RM\_LLD algorithm iterates $n_{\rm max}=5$ times to get better performance.
Thus, the computational complexity of the list RM\_LLD algorithm is 5 times larger than Algorithm 1.
Moreover, the number of transmit bits of Algorithm 1 is greater than the list RM\_LLD algorithm.


Fig. \ref{FAR} plots the false alarm rate performance obtained by Algorithm 1, Algorithm 2, and the list RM\_LLD algorithm.
The false alarm rate performance of Algorithm 2 increases slowly with the number of active devices, while the remaining algorithms increase rapidly when the number of active devices is greater than 60.
For the list RM\_LLD algorithm and Algorithm 1, since the maximum number of detected devices is set to equal to $K$, the false alarm rate plus the success rate equals to 1.
For Algorithm 2, the size of the output may not be equal to $K$.
For the setting in this paper, the size of the output in the first phase is often greater than the number of active devices.
This explains why the false alarm rate of Algorithm 2 is much worse when the number of active devicesr is small.
However, one can consider 1 minus the success rate as the false alarm rate in the second phase.


Fig.\ref{MAR} illustrates the miss rate performance of the above algorithms. Since the maximum number of detected devices in the list RM\_LLD algorithm is set to $K_{\rm max}=K$, the false alarm rate of both algorithms equals to the miss rate according to the definition in \eqref{DefiniFAR} and \eqref{DefiniMAR}. For Algorithm 2, the false alarm rate is basically larger than the miss rate, because the output size of Algorithm 2 is larger than the number of active devices for the setting in this paper.

%

Fig.\ref{CEE} describes the channel estimation error obtained by different algorithms.
Since the channel gain for different devices are generated randomly, we consider the estimation correct if the error between the estimation and the real channel coefficient is smaller than a threshold.
Still, Algorithm 2 outperforms the list RM\_LLD algorithm when the number of active devices are larger than 60.

\begin{table*}[htbp]
\renewcommand\arraystretch{1}
\centering
\caption{Comparison of CPU time (seconds) between different algorithms.}
\label{CPUtime}
\setlength{\tabcolsep}{3.5mm}{
\begin{tabular}{|l|l|l|l|l|l|l|l|l|l|}
  \hline
  $K$ & 40 & 50 & 60 & 70 & 80 & 90 & 100 & 110 & 120 \\
  \hline
  Algorithm in \cite{Hanzo} & 16.85  &  21.12  &  25.38 &  29.71  & 34.87  & 39.01  & 43.59 &  47.75 &  52.74 \\
  Algorithm 1 & 3.18 &   4.05  &  4.97  &  5.86  &  6.65  &  7.59  &  8.49  &  9.32  & 10.23  \\
  Algorithm 2 & 0.86 &	1.08 & 1.32 & 1.64 & 1.82& 2.02 & 2.21 & 2.49 & 2.63 \\
  \hline
\end{tabular}
}
\end{table*}

The CPU time (seconds) versus the number of active devices for Algorithm 2 and the list RM\_LLD algorithm is given in Table \ref{CPUtime}. The codes are executed on an Intel Core i7-5600 2.6 GHz processor.
From Table \ref{CPUtime}, it can be observed that: 1) The CPU time for all algorithms increases approximately linearly with the number of active devices; 2) The CPU time of Algorithm 1 is about 5 times smaller than that of the list RM\_LLD algorithm, which results from $n_{\rm max}=5$; 3) The CPU time of the list RM\_LLD algorithm is about $n_{\rm max}2^p=20$ times the CPU time of Algorithm 2.

We note that the number of bits can be transmitted is limited by the codeword length, and slotting further constraint the number of bits that can be transmitted. Roughly, more slots can reduce the computational complexity and increase the number of messages that can be transmitted, but simultaneously, the number of bits that can be transmitted is reduced.
Following \cite{Robert}, we use bits partitioning to increase the number of transmitting bits. Specially, each message is split into patches that can be decoded separately and then patch together.

We split the entire codeword of length $2^m$ into $2^r$ sub-blocks first, each of length $2^{p+q}$, i.e., $m=q+p+r$. Then we split each sub-block into $2^p$ slots, each of length $2^q$.
To patch together the messages transmitted in different sub-blocks, we append $l_i$ random parity check bits to patch $i,i=2,\cdots,2^r$ (The first patch is not assigned with parity check bits.).
In this case, the number of transmit bits is $B=2^r\left(\frac12q(q+3)+p-1\right)-\sum_{i=2}^{2^r}l_i$.

\begin{figure}
  \begin{center}
  \includegraphics[width=2.8in]{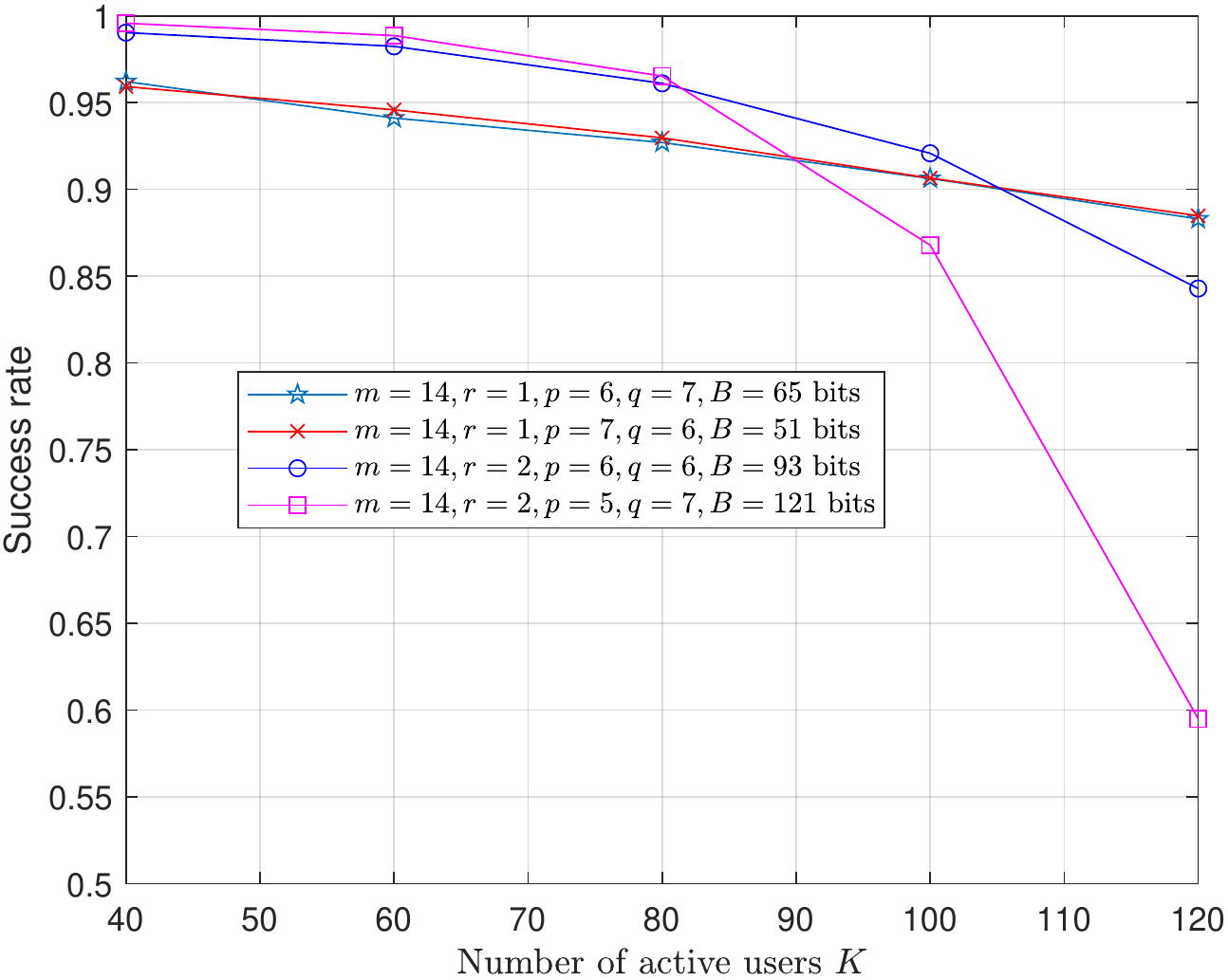}
  \end{center}
  \caption{The success rate versus the number of active devices under different number of patches and slots.}\label{patchwithoutneighbor}
\end{figure}

Fig. \ref{patchwithoutneighbor} plots the success rate versus the number of active devices under different parameter settings. We fix the codeword length to be $n=2^{14}=16,384$. When $r=1$, we set the parity bits to be $l=[0, 15]$, and $l=[0, 10, 10, 15]$ when $r=2$.
It can be observed from Fig. \ref{patchwithoutneighbor} that when the number of active devices is small, for example, $K=40-80$, a small number of slots $p=5$ is enough, and simultaneously, the success rate can be improved by increasing the length of each slot.
Moreover, with a small number of active devices, we can transmit more bits by splitting the codeword into multiple patches.
When the number of active devices is large, more slots help improve the success rate compared with increasing the length of each slot.
Besides, we can reduce the number of patches to transmit more messages.
Keeping the length of the codeword fixed, more slots allow more messages to be transmitted while simultaneously reducing the number of transmit bits $B$.
For a fix codelength $2^m$, the number of transmit bits of the list RM\_LLD algorithm in \cite{Hanzo} is also fixed, i.e., $\frac12m(m-1)$, while the number of transmit bits of Algorithm 2 is related to the choice of $q,p,r$. It is not fair to compare the list RM\_LLD algorithm and Algorithm 2 in this case. Therefore we have not drawn the curves obtained by the list RM\_LLD algorithm in Fig. \ref{patchwithoutneighbor}.

\subsection{Performance Comparison with Out-of-the-cell Devices}
In this section, we further assume that there are interfering transmissions from out-of-the-cell devices.
We fix the codeword length to be $2^{12}=4,096$ and we consider a network with $K$ active devices uniformly distributed in a $500\times 500\ {\rm m^2}$ square.
According to \eqref{averageneighbor}, if $K=1000$ the average number of in-cell devices of the AP is $K^{*}\approx11$ , and if $K=8000, K^{*}\approx90$.
Likewise, we assume the list RM\_LLD algorithm in \cite{Hanzo} and Algorithm 1 know the number of active devices $K^{*}$ when decoding, while Algorithm 2 does not.

\begin{figure*}[htbp]
\centering

\subfigure[]{
\begin{minipage}[t]{0.45\linewidth}
\centering
\includegraphics[width=2.2in]{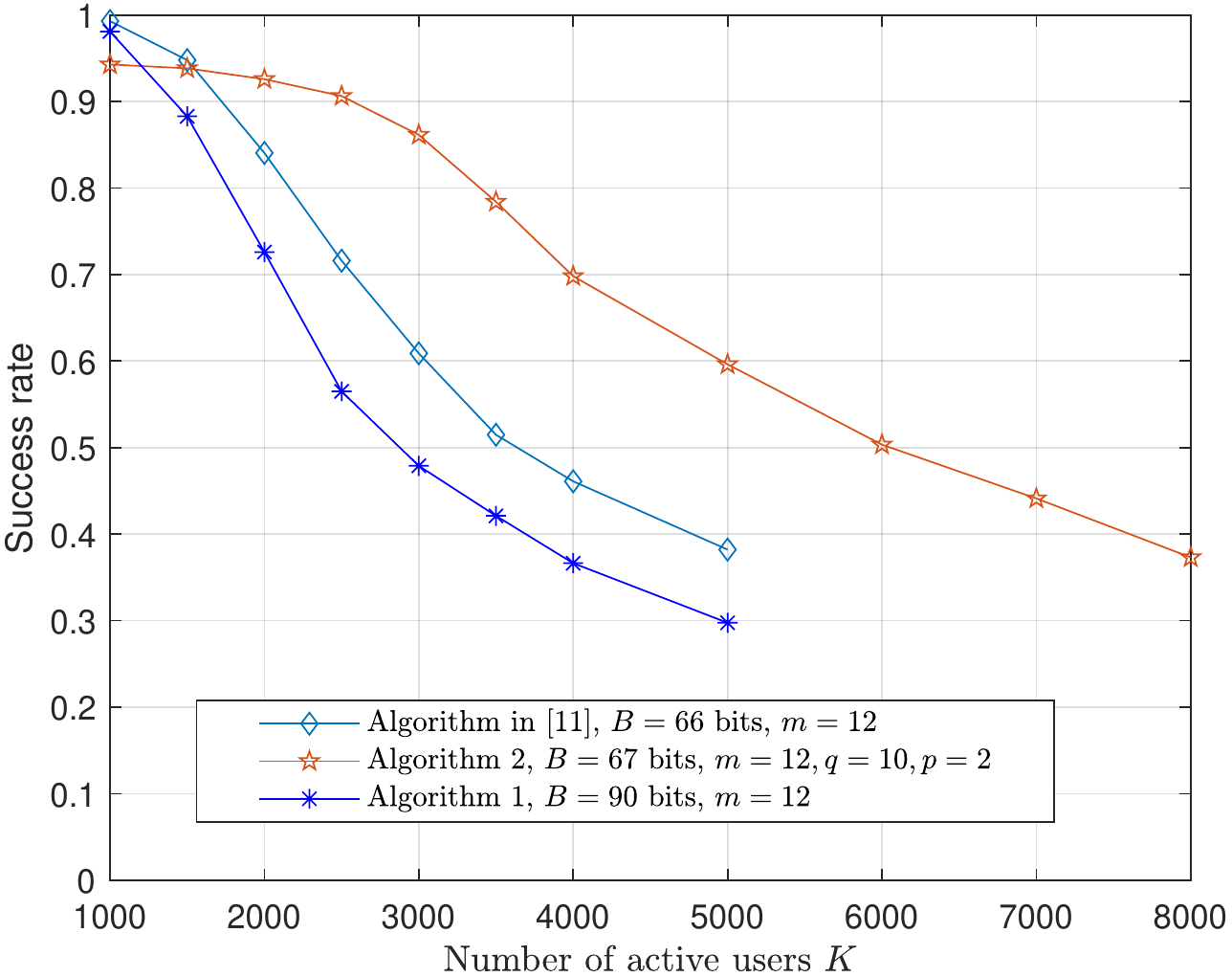}
\end{minipage}%
\label{SDPneighbor}
}%
\subfigure[]{
\begin{minipage}[t]{0.45\linewidth}
\centering
\includegraphics[width=2.2in]{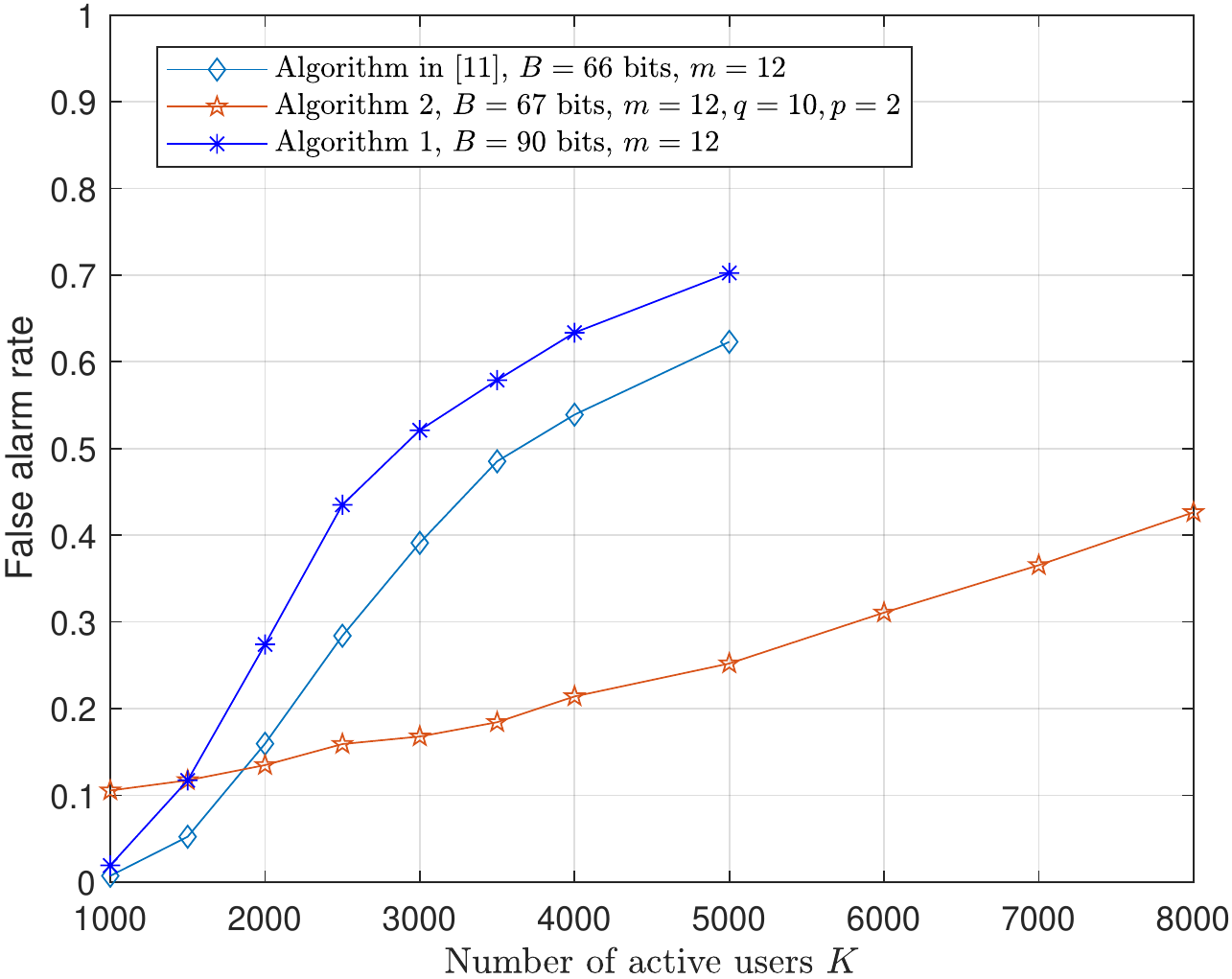}
\end{minipage}%
\label{FARneighbor}
}%

\subfigure[]{
\begin{minipage}[t]{0.45\linewidth}
\centering
\includegraphics[width=2.2in]{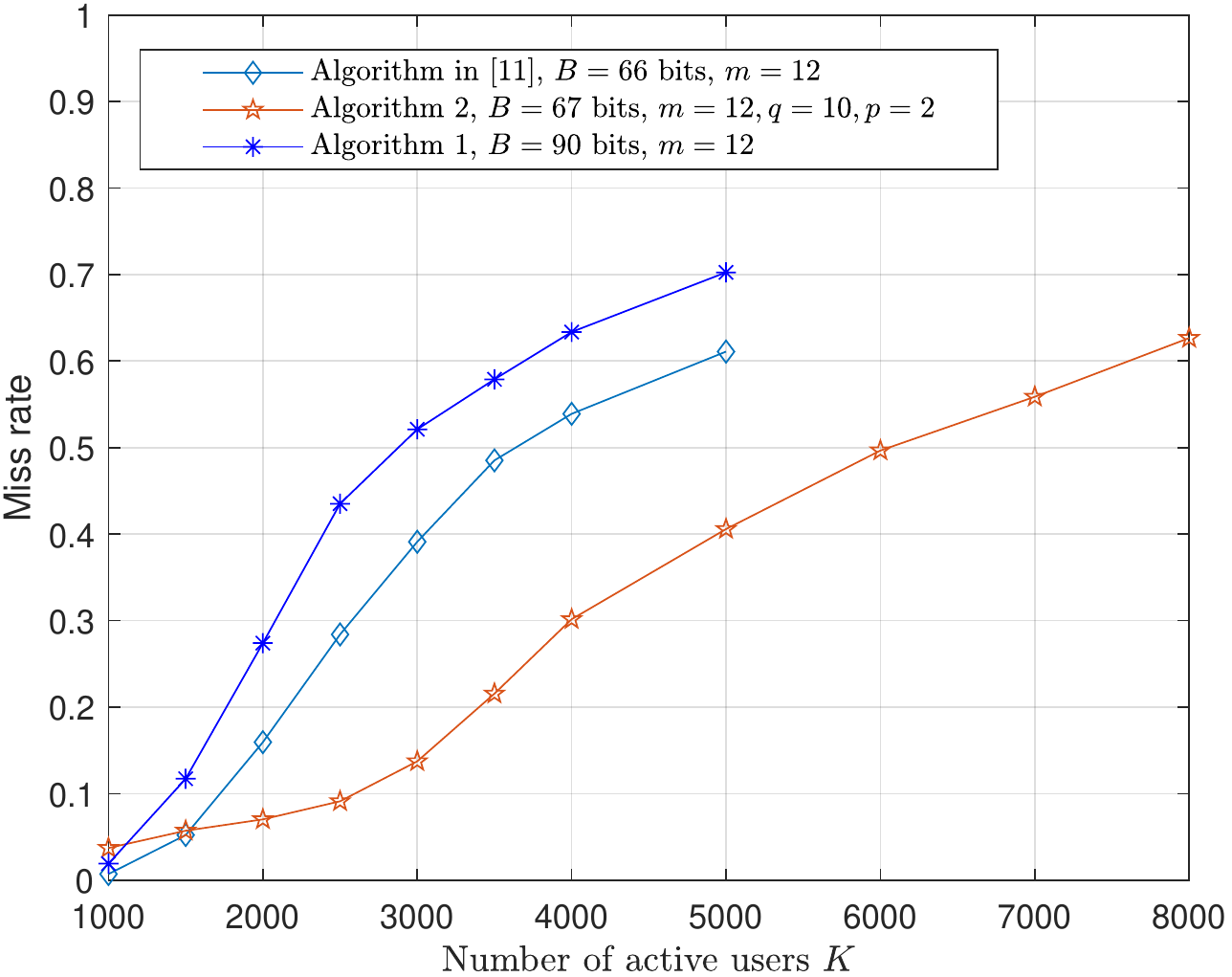}
\end{minipage}
\label{MARneighbor}
}%
\subfigure[]{
\begin{minipage}[t]{0.45\linewidth}
\centering
\includegraphics[width=2.2in]{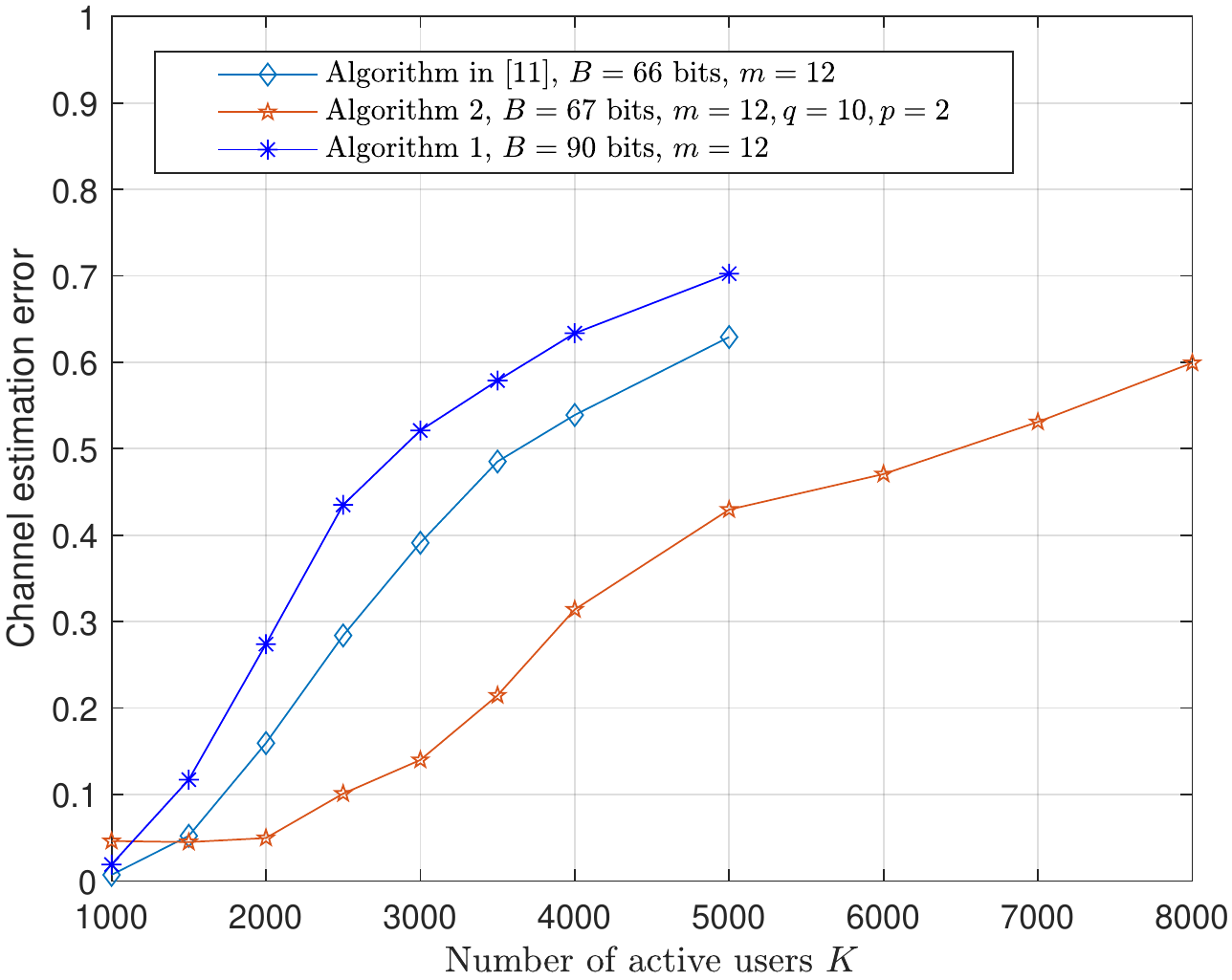}
\end{minipage}
\label{CEEneighbor}
}%

\centering
\caption{Performance comparison between different RM decoding algorithms with varying number of active devices in the $500\times 500\ {\rm m}^2$ area. (a) The success rate versus $K$, (b) the false alarm rate versus $K$, (c) the miss rate versus $K$, and (d) the channel estimation error versus $K$.}
\label{neighbor}
\end{figure*}

Fig. \ref{neighbor} depicts the success rate, false alarm rate, miss rate, and channel estimation error obtained by Algorithm 2 and the algorithm in \cite{Hanzo}, respectively.
It can be observed that Algorithm 2 outperforms the list RM\_LLD algorithm when $K>1500$.
In addition, the performance of Algorithm 2 drops gracefully as the number of active devices increases, while the performance of Algorithm 1 and the list RM\_LLD algorithm drop rapidly with $K$.
Similar to Fig. \ref{FAR}, Fig. \ref{FARneighbor} shows that the false alarm rate of Algorithm 2 is much worse than the list RM\_LLD algorithm when the number of active devices is small. This is because the size of the output of Algorithm 2 is larger than the number of active devices in the first phase under the setting used in this paper. One can consider 1 minus the success rate as the false alarm rate in the second phase.

\begin{figure}
  \begin{center}
  \includegraphics[width=2.5in]{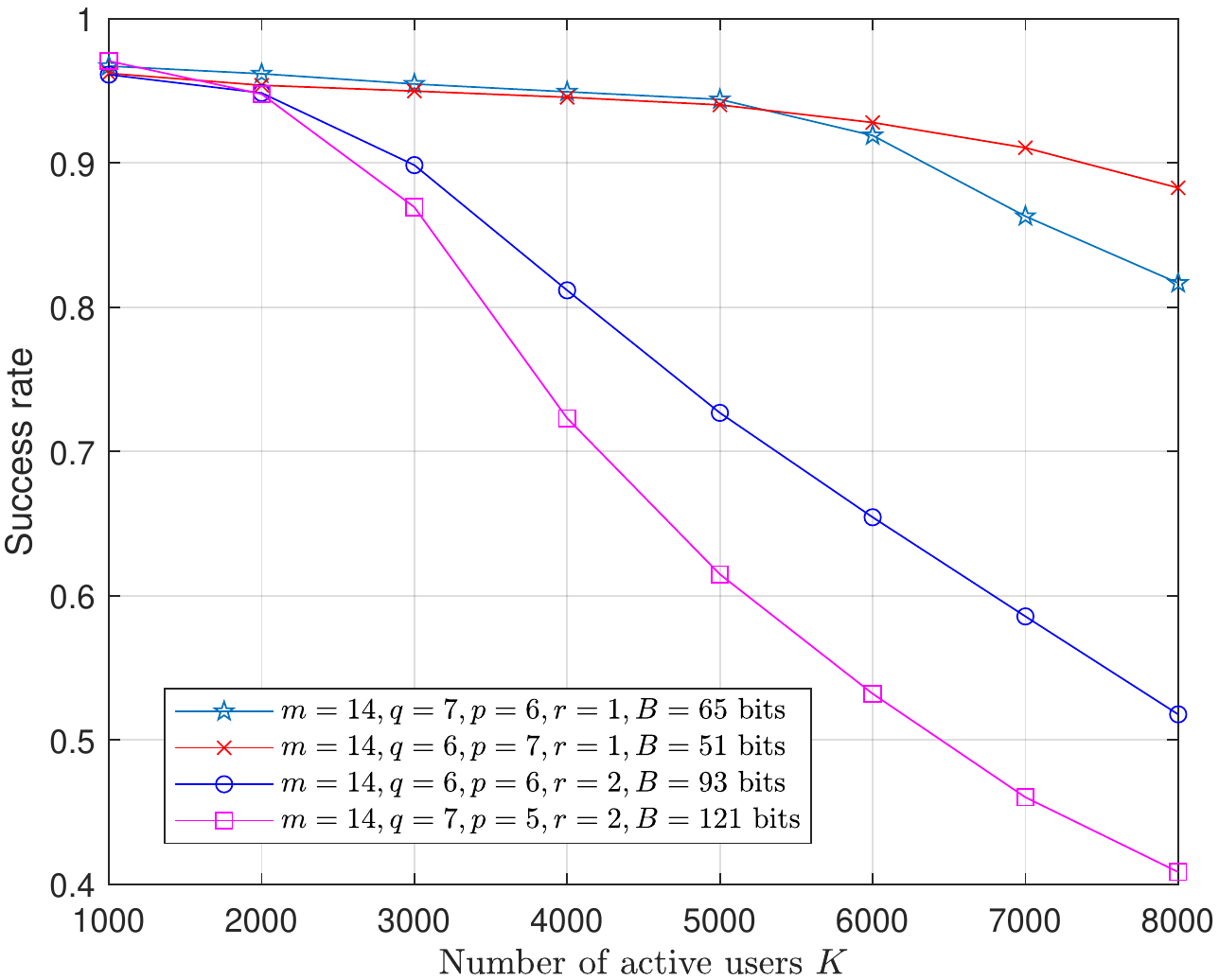}
  \end{center}
  \caption{The success rate versus the number of active devices under different number of patches and slots.}\label{patchwithneighbor}
\end{figure}

Fig. \ref{patchwithneighbor} plots the success rate versus the number of active devices under different parameter settings. We use the same setting as in Fig. \ref{patchwithoutneighbor}.
Likewise, when the number of active devices is small, the success rate can be improved by increasing the length of each slot and we can transmit more bits by splitting the codeword into multiple patches.
When the number of active devices is large, more slots help improve the success rate compared with increasing the length of each slot.

%
%
%

\section{Conclusion}\label{conclusion}
This paper builds on some state-of-art coding schemes to develop a new scheme with efficient algorithms for massive access in presence of channel uncertainties.
We have presented two fast RM decoding algorithms, i.e., Algorithm 1 and Algorithm 2, for active device detection and channel estimation for massive access.
Both algorithms are based on a new relationship between the RM sequences and its subsequences.
Specifically, Algorithm 2 is an enhanced RM decoding algorithm by applying slotting and message passing on Algorithm 1.
The complexity of Algorithm 2 is significantly lower than Algorithm 1 and the list RM\_LLD algorithm given in \cite{Hanzo}, which makes Algorithm 2 an attractive random access scheme for practice. Simulation results demonstrate that Algorithm 2 outperforms Algorithm 1 and another recent scheme, especially when the number of active devices is large.


\end{document}